\documentclass[a4paper,fleqn]{cas-dc}

\usepackage[authoryear]{natbib}

\usepackage[english]{babel} 
\usepackage{siunitx} 
\DeclareSIUnit{\sliter}{sL}

\usepackage{xfrac,nicefrac} 
\usepackage{enumitem} 
\usepackage{subfig} 

\usepackage{placeins} 
\usepackage{float} 
\usepackage{varioref}

\usepackage{mathtools}
\usepackage[makeroom]{cancel}

\usepackage{amssymb,amsmath,amsthm}
\newtheorem*{remark}{Remark}

\usepackage{algorithm}
\usepackage[noend]{algpseudocode}
\usepackage{algcompatible}

\def\tsc#1{\csdef{#1}{\textsc{\lowercase{#1}}\xspace}}
\tsc{APC}    
\tsc{ASROPA} 
\tsc{DRTO}   
\tsc{EKF}    
\tsc{EMPC}   
\tsc{HRTO}   
\tsc{IFAC}   
\tsc{MHE}    
\tsc{MILP}   
\tsc{MINLP}  
\tsc{ROPA}   
\tsc{RTO}    
\tsc{SS}     
\tsc{SSD}    
\tsc{SSRTO}  
\tsc{UKF}

\tsc{DAE}
\tsc{ODE}

\begin{document}
\let\WriteBookmarks\relax
\def\floatpagepagefraction{1}
\def\textpagefraction{.001}
\shorttitle{RTO using transient measurements on experimental rig}
\shortauthors{Matias et~al.}

\title [mode = title]{Steady-state Real-time Optimization Using Transient Measurements on an Experimental Rig}       \tnotemark[1]
\tnotetext[1]{The authors acknowledge financial support from the Norwegian Research Council/Intpart, SUBPRO, grent number: 237893. Galo A. C. Le Roux acknowledges CNPq, the National Council for Scientific and Technological Development for his productivity grant 312049/2018-8.}

\author[1]{Jos\'e Matias}[orcid=0000-0002-1094-6145]
\credit{Conceptualization, Methodology, Coding Implementation, Formal analysis, Data Curation, Writing - Original Draft}

\author[1,2]{ Julio P. C. Oliveira}
\credit{Conceptualization, Methodology, Coding Implementation, Writing - Original Draft}

\author[2]{ Galo A.C. Le{ }Roux}
\credit{Conceptualization, Writing - Review \& Editing, Supervision, Project administration, Funding acquisition}

\author[1]{ Johannes J\"aschke}
\cormark[1]
\ead{johannes.jaschke@ntnu.no}
\credit{Conceptualization, Writing - Review \& Editing, Supervision, Project administration, Funding acquisition}

\address[1]{Chemical Engineering Department, Norwegian University of Science and Technology, Sem S{\ae}landsvei $4$, Kjemiblokk $5$, Trondheim, Norway}

\address[2]{Chemical Engineering Department, University of S\~ao Paulo, Av. Prof. Lineu Prestes, $580$, Conjunto das Químicas/bloco $18$, S\~ao Paulo, Brazil}

\cortext[cor1]{Corresponding author}

\begin{abstract}
Real-time optimization with persistent parameter adaptation (\ROPA) is an \RTO approach, where the steady-state model parameters are updated dynamically using transient measurements. Consequently, we avoid waiting for a steady-state before triggering the optimization cycle, and the steady-state economic optimization can be scheduled at any desired rate. The steady-state wait has been recognized as a fundamental limitation of the traditional \RTO approach. In this paper, we implement \ROPA on an experimental rig that emulates a subsea oil well network. For comparison, we also implement traditional and dynamic \RTO. The experimental results confirm the \textit{in-silico} findings that \ROPA's performance is similar to dynamic \RTO's performance with a much lower computational cost. Finally, we present some guidelines for \ROPA's practical implementation. 
\end{abstract}

\begin{graphicalabstract}
\includegraphics{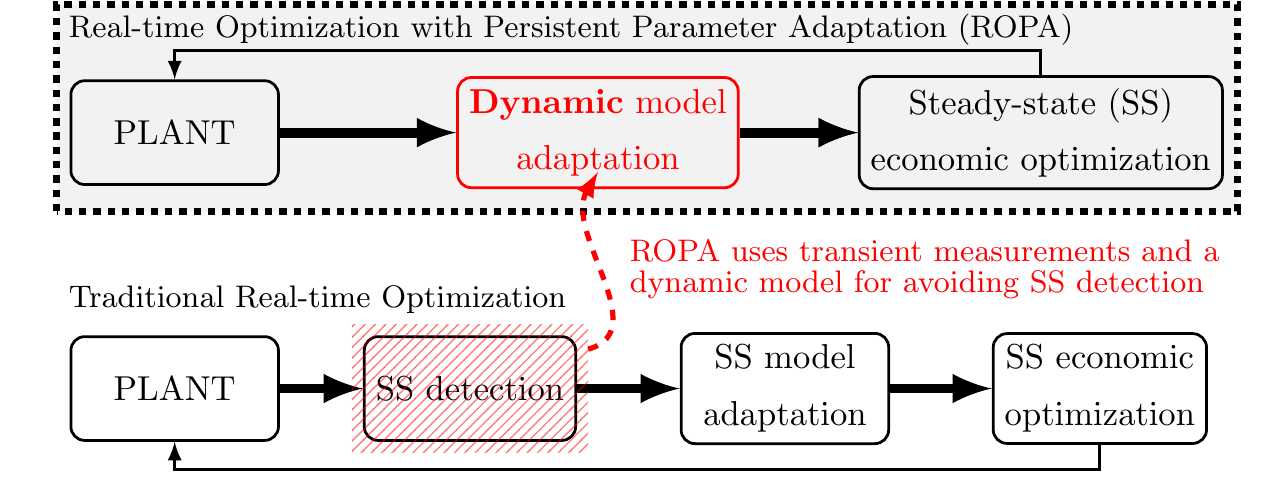}
\end{graphicalabstract}

\begin{highlights}

\item \RTO with persistent parameter adaptation (\ROPA), traditional, and dynamic \RTO (\DRTO) are implemented on lab scale plant;
\item \ROPA avoided \SS detection and showed economic performance comparable to \DRTO.
\item Guidelines for the practical implementation of \ROPA

\end{highlights}


\maketitle

\section{Introduction}

\begin{figure*}
	\centering
	\subfloat[\small Steady-state \RTO]{
    \includegraphics[trim= 0.5cm 0cm 0cm 0cm, clip=true, width=0.28\textwidth]{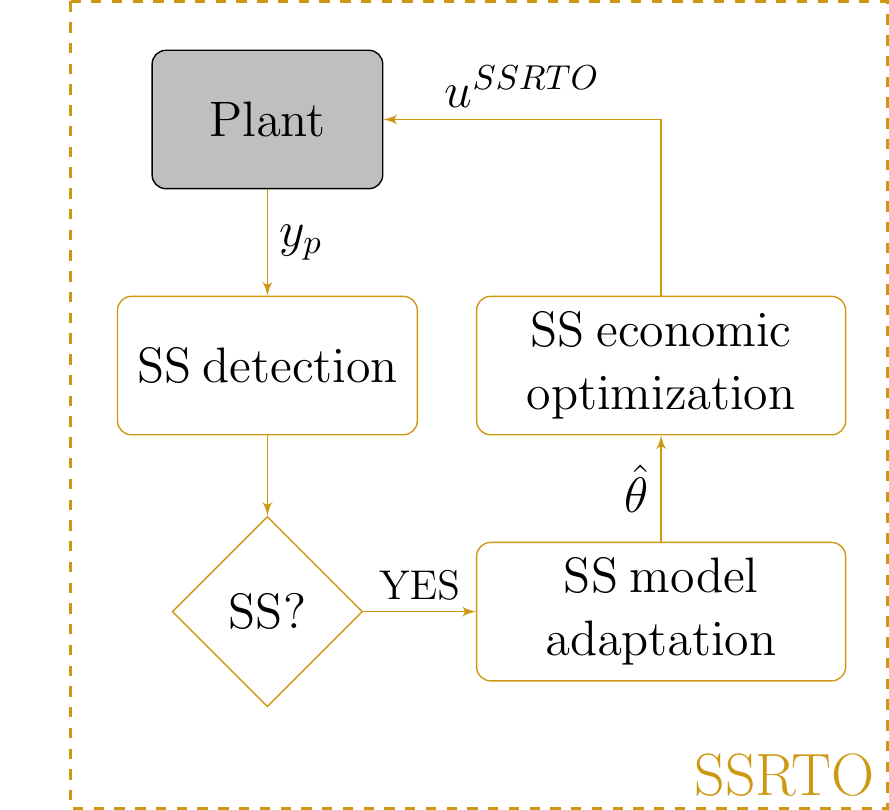}\label{fig:BD_SSRTO}}
    \subfloat[\small{Dynamic \RTO}]{
    \includegraphics[trim= 0.5cm 0cm 0cm 0cm, clip=true, width=0.286\textwidth]{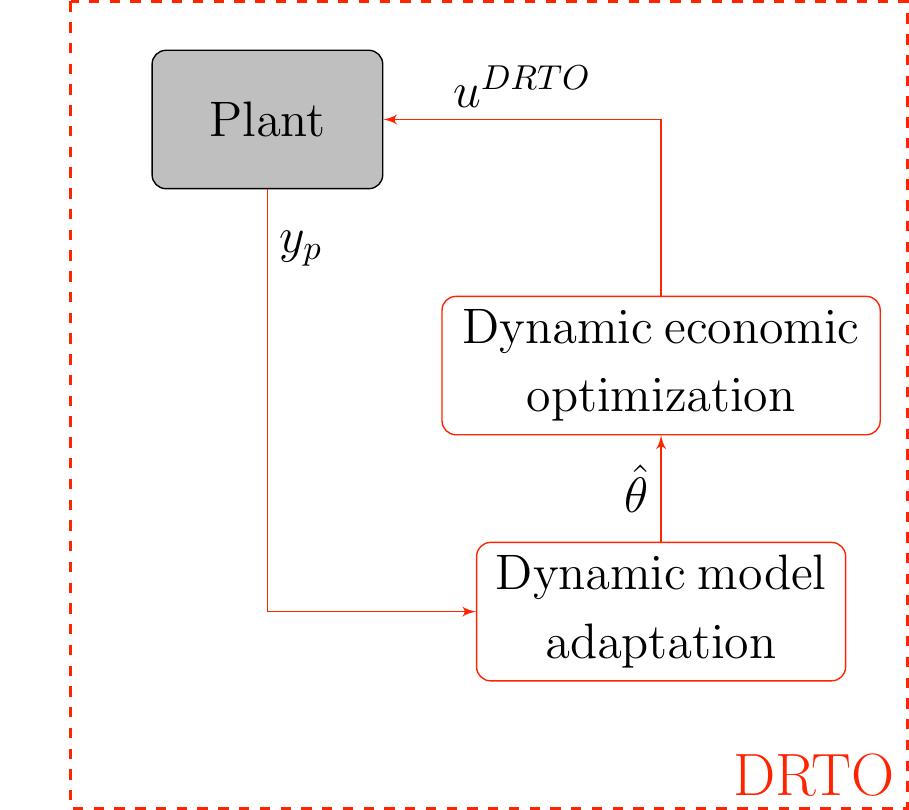}\label{fig:BD_DRTO}}
    \subfloat[\small{\RTO with persistent parameter adaptation}]{
    \includegraphics[trim= 0.5cm 0cm 0cm 0cm, clip=true, width=0.28\textwidth]{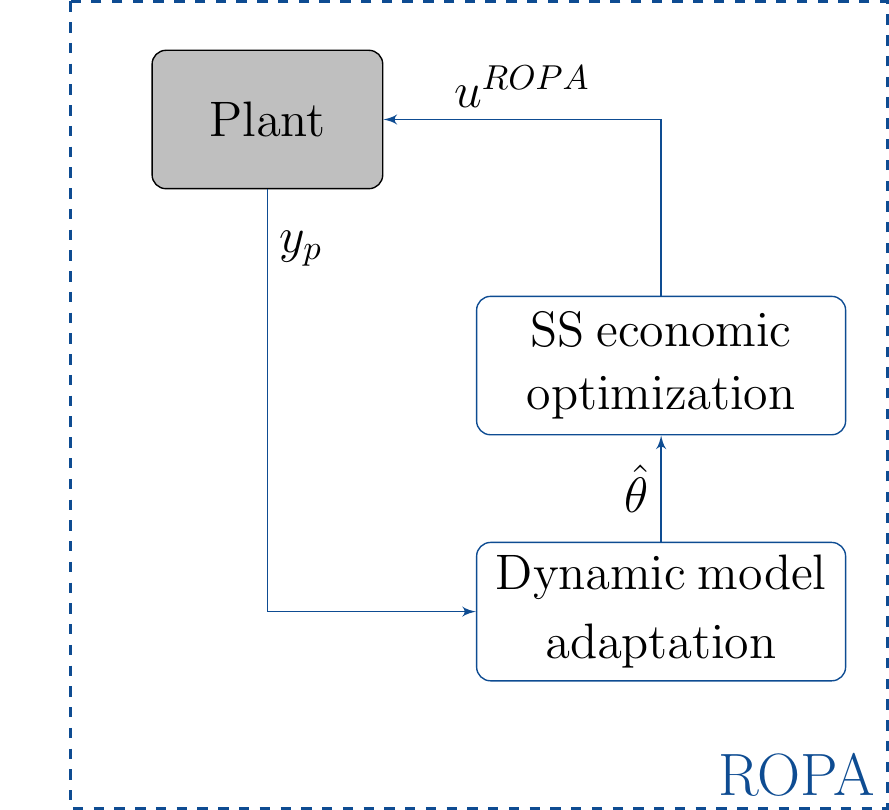}\label{fig:BD_ROPA}}
	\caption{Block diagram comparing the three approaches. Here, $u$ are the computed inputs (system manipulated variables); $\hat{\theta}$ the estimated parameters, and $y_p$ the plant measurements.}
	\label{fig:blockComparison}
\end{figure*}

Real-time Optimization (\RTO) is a production optimization technique that aims at improving plant economic performance in real-time. In the traditional steady-state \RTO (\SSRTO), which was originally proposed by \cite{chen:joseph:1987}, a rigorous steady-state model is adapted to the current plant state. Next, the updated model is used for computing the optimal operating point, which is then implemented in the plant (see Figure~\ref{fig:BD_SSRTO}). \cite{cutler:perry:1983} found that combining \SSRTO with advanced process control can increase plant profit up to \SI{10}{\percent}. Despite its potential benefits, the traditional \RTO is still not widely used in practice \citep{darby:nikolaou:jones:nicholson:2011}.

Multiple challenges and technical issues are associated with this reluctant acceptance (e.g. corrupted information coming from sensors, plant-model mismatch, interface between \RTO and advanced control, etc.). Among them, the need to wait for a steady-state (\SS) before triggering the optimization cycle has been recognized as a fundamental limitation \citep{friedman:1995}. This drawback comes from the fact a  \emph{steady-state} model of the plant is used for finding the optimal operating strategy. Since it uses a static model, the system \emph{must} be at steady-state for a reliable update of the model parameters. Otherwise, the computed operating conditions are likely to be sub-optimal and potentially hazardous to the plant \citep{engell:2007}. 

\begin{figure}[ht]
	\centering
	\includegraphics[width=0.38\textwidth]{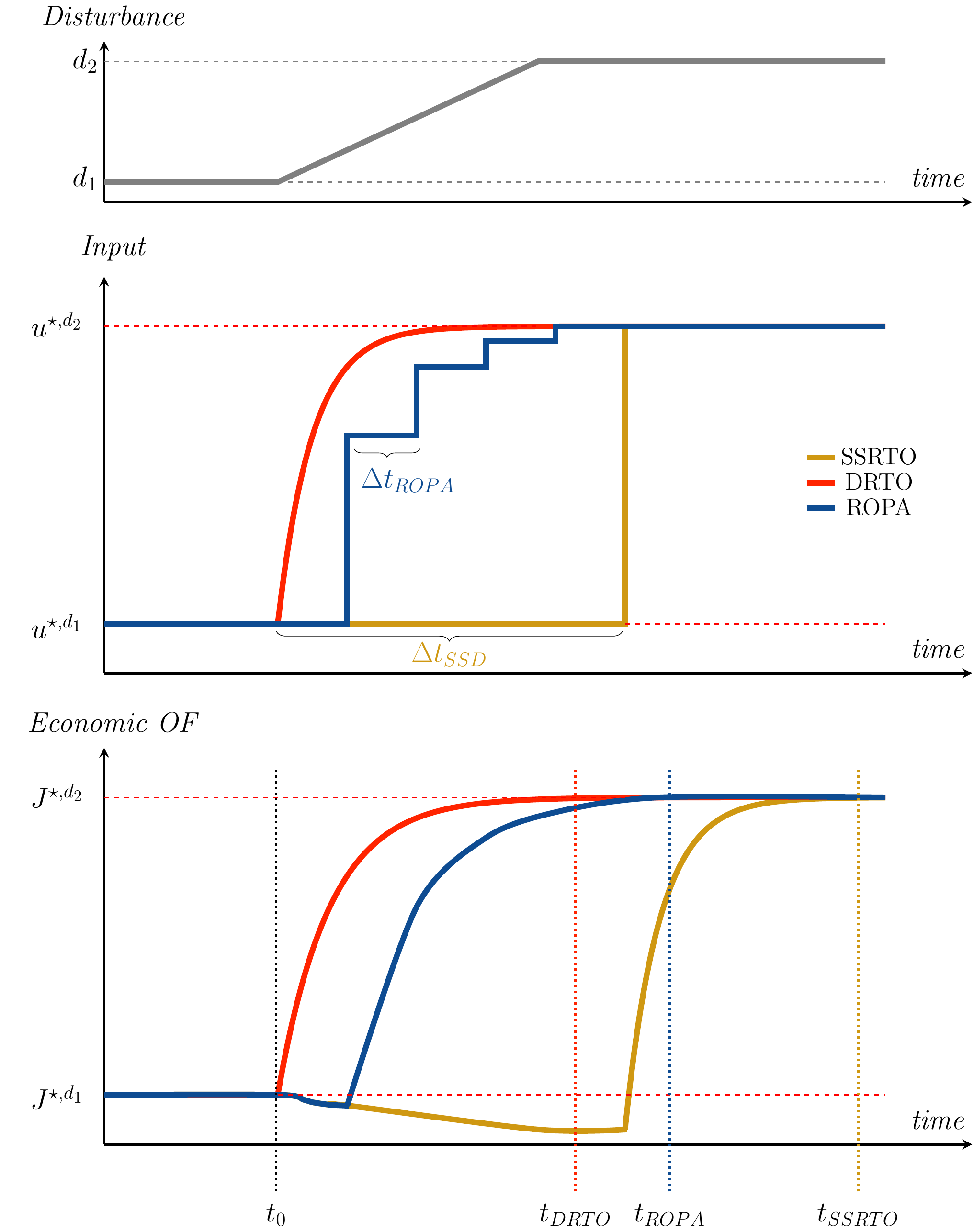}
	\caption{Idealized response of the three approaches to a ramp disturbance $(d_1 \to d_2)$. The computed inputs $u^\star$ and the resulting economic index $J$ are shown. $t_{SSRTO}$, $t_{DRTO}$ and $t_{ROPA}$ represent the time the approaches take to drive the plant to the new steady-state optimum $J^{\star,d_2}$. $\Delta t_{SSD}$ is the steady-state wait time of the traditional \RTO and $\Delta t_{ROPA}$ is \ROPA's execution rate.}
	\label{fig:typicalProfiles}
\end{figure}

Figure~\ref{fig:typicalProfiles} illustrates this limitation, showing the schematic response of the traditional \RTO to a ramp disturbance. At $t_0$, the disturbance enters the system. The economic optimization executes only at $t_0 + \Delta t_{SSD}$. This delay is caused by a combination of the time required for the \SSD procedure to detect steady-state and the process settling time, which can be considerably long for persistent disturbance as the one shown in Figure~\ref{fig:typicalProfiles}. As an additional practical limitation, identifying if the data comes from a stationary or transient period is challenging \citep{menezes:2016}. Practitioners tend to be conservative and only few data periods are identified as steady-state \citep{camara:quelhas:pinto:2016}. As a result, \RTO is too seldom executed, decreasing its economic benefits. 

The issue of steady-state wait for updating the steady-state model can be addressed by using a dynamic real-time optimization (\DRTO) approach (Figure~\ref{fig:BD_DRTO}). Although \DRTO is conceptually similar to \SSRTO (i.e. a model adaptation step followed by an economic optimization step), it relies fully on a dynamic model of the plant. Thus, it is possible to use transient measurements in the adaptation step, avoiding the steady-state wait. In Figure~\ref{fig:typicalProfiles}, we schematically compare \DRTO and \SSRTO. Since we do not have to wait for a new steady-state, \DRTO drives the plant to the new optimum immediately after the disturbance starts, reaching the new optimum faster than \SSRTO. Despite being conceptually attractive, only a few real-world applications were published (for example, \cite{rawlings:patel:etal:2018economic}) and, to the best of the authors' knowledge, there are no mature \DRTO industrial implementations reported in the literature. 
Alternatively, a hybrid approach (Figure~\ref{fig:BD_ROPA}) can be used, where we use transient measurements and a dynamic model of the system for the adaptation step (like in \DRTO), whereas the economic optimization is performed using a steady-state model (like in \SSRTO). As a consequence, the steady-state economic optimization can be scheduled at any arbitrary rate avoiding the need to wait for a  steady-state. This scheme was independently proposed by \cite{krishnamoorthy:foss:skogestad:2018} using the name Hybrid RTO (\HRTO), and by \cite{matias:leroux:2018}, who called it Real-time Optimization with Persistent Parameter Adaptation (\ROPA). We will use the latter nomenclature in this paper. 

In this paper, we implement the three approaches mentioned above on an experimental rig that emulates a subsea oil well network and compare the results. An important contribution of this paper is to confirm the previous \textit{in-silico} findings: since \ROPA avoids the steady-state wait, the optimization frequency increases, improving the overall economic performance to levels similar to \DRTO with a much lower computational effort. Along with using experimental evidence to support \ROPA's capabilities, we also propose some guidelines to its practical implementation. Our goal is to support practitioners with important model design decisions and also provide help for choosing values for \ROPA's tuning parameters. 

\section{Real-time optimization with persistent parameter adaptation - ROPA}

\ROPA works on the assumption that the \SS predictions of a sufficiently accurate dynamic model are a good indication of the future plant steady-state. Then, by continuously estimating the model parameters, we capture the effect of the current disturbances in the plant steady-state that is yet to be achieved, i.e. current disturbances are measured and can be accounted for before their effect is fully realized in the system. Consequently, if we continuously adapt the value of inputs $u$ based on this future steady-state, \ROPA can drive the system to the desired stationary optimum. 

Note that, since \SSD is not necessary, \ROPA can trigger the model adaptation and update the inputs $u$ every $\Delta t_{ROPA}$, which is the approach's sampling time  (Figure~\ref{fig:typicalProfiles}). It must be chosen such that relevant process dynamics and disturbances are captured. For a discussion on how to tune this parameter, refer to Section~\ref{sec:ropatuning}. 

As a consequence of continually updating the input $u$, \ROPA reaches the new optimal steady-state point $J^{\star,d_2}$ faster than \SSRTO. And, in the \textit{worst-case} scenario, its performance is \textit{at least as} good as \SSRTO. When compared to \DRTO, the input sequence computed by \ROPA is sub-optimal during the transients because they are not optimized. 

\subsection{Why using a SS model instead of a dynamic for economic optimization?}
Despite having sub-optimal transients, there are advantages of running a steady-state instead of a dynamic economic optimization. As mentioned, one of the main challenges with \DRTO is the large amount of computation required. Both the dynamic model adaptation and dynamic economic optimization problems need to be solved every sampling time, which may result in a computation delay and subsequently optimization performance degradation as well as system instabilities \citep{findeisen:allgower:2004}. 

On the other hand, solving only the dynamic model adaptation problem (which, in addition, can be carried out by a recursive method) is much more attractive from a computation cost point of view. Especially, if this alternative leads to similar economic performances. For example, \cite{krishnamoorthy:foss:skogestad:2018} applied \ROPA and \DRTO on a simulated subsea well network and showed that both approaches achieved very similar economic performance while \ROPA had much lower computational requirements. 

Another important advantage of using steady-state models is related to solving mixed-integer (non)linear programming (\MILP/\MINLP) optimization problems. For example, in subsea oil well networks, the well flow can be routed to one of different flowlines \citep{foss:knudsen:grimstad:2018}. If the routing becomes a decision variable of a dynamic economic optimization, the problem needs to be formulated as a dynamic \MILP/\MINLP, which is much more challenging to solve than the steady-state version. 

In addition, \ROPA has been extended to a plant-wide optimization context using an asynchronous parameter adaptation strategy \citep{matias:leroux:2020}. Here, only a subset of the plant-wide steady-state model parameters is adapted dynamically. This strategy allows the \emph{plant-wide} optimization cycle to be triggered much more frequently. From an economic point of view, it is advantageous to solve the plant-wide economic problems since the trade-offs between plant units are taken into account \citep{friedman:1995}. Moreover, the complete model is solved and no decomposition techniques are required. 

\subsection{ROPA building blocks}

According to Figure~\ref{fig:blockComparison}, the two main building blocks of \ROPA are \textit{Dynamic model adaptation} and \textit{\SS economic optimization}.

\subsubsection{Dynamic model adaptation}\label{subsec:DynModelAdapt}
In \SSRTO, the model adaptation is typically  formulated as a parameter estimation problem \citep{darby:nikolaou:jones:nicholson:2011}, in which the model parameters (or a subset of them) are updated such that the difference between model predictions and averaged plant
measurements is minimized (see Section~\ref{sec:SSRTO_DRTO} for more details). In contrast, there are different strategies based on different paradigms for dynamic model adaptation. For instance: optimization-based methods, like moving horizon estimator \citep{rao:rawlings:2000}, and recursive methods, such as recursive least squares estimation, extended Kalman filter (\EKF), and unscented Kalman Filter (\UKF). 

\paragraph{Extended Kalman filter (EKF):}\label{sec:ekfformulation}
this is the most common method used for nonlinear dynamic state/parameter adaptation in practice \citep{schneider:christos:2013}. For implementing it, we model the system as:
\begin{equation}\label{eq:DynModel}
    \begin{aligned}
        \boldsymbol{\dot{x}}(t) &= \boldsymbol{f}(\boldsymbol{x}(t),\boldsymbol{u}(t) \ | \boldsymbol{\theta}(t)) + \boldsymbol{\upsilon}(t) \\
        \boldsymbol{y}(t) &= \boldsymbol{h}(\boldsymbol{x}(t),\boldsymbol{u}(t)) + \boldsymbol{\omega}(t)
    \end{aligned}
\end{equation}

\noindent where, $\boldsymbol{x}$ are the model states, $\boldsymbol{u}$ the set of inputs (manipulated variables), $\boldsymbol{y}$ the model predictions, and $\boldsymbol{\theta}$ the model parameters. The variables $\boldsymbol{\upsilon}$ and $\boldsymbol{\omega}$ are the process and measurement noise, both are assumed to be uncorrelated zero-mean Gaussian random processes. The function $\boldsymbol{f}$ is the system dynamic model\footnote{For simplicity, we represent the model as a system of ordinary differential equations (\ODE) but it could be easily adapted to a system of differential-algebraic equations (\DAE).}, and $\boldsymbol{h}$ is a function mapping $\boldsymbol{x}$ and $\boldsymbol{u}$ to the model outputs $\boldsymbol{y}$.

\EKF is implemented by first using Equation~\ref{eq:DynModel} for evolving the current state estimates in time. Next, $\boldsymbol{f}$ and $\boldsymbol{h}$ are linearized using an appropriate method. Then, since the nonlinear model in Equation~\ref{eq:DynModel} is now approximated by a linear model, the estimates and their covariance matrix can be updated using the standard Kalman filter equations \citep{walter:pronzato:1997}. 

Since in \ROPA context, we are also interested in estimating the parameters, we need to adapt our model in Equation~\ref{eq:DynModel}. First, we assume that the parameter ``dynamics'' follow a Gaussian random walk model. Then, we extend $\boldsymbol{f}$ with the parameter ``dynamics'' as well as the state vector to include $\boldsymbol{\theta}$. As a result, we can use \EKF's framework for obtaining an estimation of the parameter $\boldsymbol{\hat{\theta}}$. For more details, refer to e.g. \cite{walter:pronzato:1997}.

\subsubsection{Steady-state economic optimization}\label{subsec:SSeconOpt}
The economic optimization is executed after the model is adapted at every $\Delta t_{ROPA}$. In this step, new inputs $\boldsymbol{u}$ are calculated such that an economic criterion $J$ is optimized. $J$ is typically chosen as profit (i.e. product value – feed costs – variable costs). The optimized inputs are then implemented in the plant. The \SS economic optimization problem can be posed as:
\begin{equation}\label{eq:ssopt}
    \begin{aligned}
        & \boldsymbol{u^\star}, \boldsymbol{x^\star} = & & \underset{\boldsymbol{x}, \boldsymbol{u}}{\arg \max} \quad 
        J(\boldsymbol{y},\boldsymbol{u}) \\
        & \text{s.t.}  & & 0 = \boldsymbol{f}(\boldsymbol{x},\boldsymbol{u} \ | \boldsymbol{\hat{\theta}}) \\
        & & &\boldsymbol{y} = \boldsymbol{h}(\boldsymbol{x},\boldsymbol{u}) \\
        & & &\boldsymbol{g}(\boldsymbol{y},\boldsymbol{u}) \leq 0
    \end{aligned}
\end{equation}

\noindent where, $\boldsymbol{g}$ is the set of operational constraints and $\boldsymbol{\hat{\theta}}$, the current parameter estimates from the dynamic model adaptation step. 

Note that we assume that parameters values are estimated such that the process model is a proper representation of the plant, i.e. no plant-model mismatch. Therefore, we do not consider the noise $\boldsymbol{\upsilon}$ and $\boldsymbol{\omega}$ models in the economic optimization. If that does not apply, production optimization techniques that cope with plant-model mismatch (e.g. \cite{marchetti:francois:faulwasser:bonvin:2016}), robust optimization schemes (e.g. \cite{golshan:etal:2008}), or online model updated schemes (e.g. \cite{matias:jaschke:2021}) can be used. 

\section{\SSRTO and \DRTO implementation}\label{sec:SSRTO_DRTO}

The main goal of the paper is to compare the performance of \ROPA with the state-of-the-art production optimization approaches (steady-state and dynamic \RTO) in a physical system. In the next subsections, we briefly present both approaches and discuss their main characteristics. For a more extensive review of \SSRTO, refer to \citep{darby:nikolaou:jones:nicholson:2011}. For \DRTO, see \citep{srinivasan:palanki:bonvin:2003}. Despite being focused on batch processes, this paper has a complete description of \DRTO building blocks and solution methods. 

\subsection{Steady-state \RTO}\label{sec:SSRTOshort}
\subsubsection{SS Detection}
The first step of the \SSRTO implementation is the steady-state detection (Figure~\ref{fig:BD_SSRTO}). There are many different \SSD procedures available in the literature. Typically, they compare intervals of measurements using statistical properties, such as hypothesis tests using F-statistic \citep{alekman:1994}, Student t-test \citep{kelly:hedengren:2013}, and R-statistic combined with first-order filters for the measurements \citep{cao:rhinehart:1995}. Although the methods rely on statistical theory, they strongly depend on tuning inputs, such as filter gains and tolerances \citep{camara:quelhas:pinto:2016}. Therefore, it is common that different procedures applied to the same data set yield different results \citep{menezes:2016}. Independently of the chosen procedure, high-frequency or persistent disturbances can hinder the start of the optimization cycle. Also, if the system to be optimized is composed of several units, \SSRTO is triggered only if \emph{all} units are at \SS, which significantly decreases the number of optimization runs \citep{matias:leroux:2020}. In our implementation, this step is carried out using a Student t-test described in details in Appendix~\ref{sec:appendix}.

\subsubsection{Model adaptation and economic optimization}
After detecting the steady-state, the \SS model adaptation step is triggered (see Figure~\ref{fig:blockComparison}). The goal is to find the values of $\boldsymbol{\theta}$ that minimize the difference between the \SS model predictions $\boldsymbol{y}$ and the current plant measurements $\boldsymbol{y_{p}}$. The model adaptation problem is posed as: 
\begin{equation}\label{eq:ssmpa}
    \begin{aligned}
        & \boldsymbol{\hat{\theta}}, \ \boldsymbol{\hat{x}} = & & \underset{\boldsymbol{x},\boldsymbol{\theta}}{\arg \min} \quad
        || \boldsymbol{y_{p}} - \boldsymbol{y}||_V \\
         & \text{s.t.}  & & 0 = \boldsymbol{f}(\boldsymbol{x},\boldsymbol{u_p} \ | \boldsymbol{\theta}) \\
        & & &\boldsymbol{y} = \boldsymbol{h}(\boldsymbol{x},\boldsymbol{u_p}) \\
        & & & \boldsymbol{\theta} \in \boldsymbol{\Theta}
    \end{aligned}
\end{equation}

\noindent where, $\boldsymbol{\Theta}$ is the allowable parameter constraint set; and $|| \cdot ||$ is a norm weighted by a matrix $V$, which is usually chosen as the identity matrix or inverse of the covariance of the measurements $\boldsymbol{y_p}$. $\boldsymbol{u_p}$ are the inputs currently on the plant. After adapting the parameters, the \SS economic optimization uses the updated steady-state model to find $\boldsymbol{u}^{SSRTO}$. This step is performed by solving the optimization problem given by Equation~\eqref{eq:ssopt} (as in \ROPA).

\subsection{Dynamic \RTO}
The dynamic parameter estimation step is carried out in the same way as for \ROPA. We also use \EKF in the dynamic \RTO implementation, but any other dynamic estimator is applicable.

\subsubsection{Dynamic Economic Optimization}
The dynamic economic optimization problem is defined on a prediction horizon $N_p$. It is stated as follows:

\begin{equation}\label{eq:dynopt}
    \begin{aligned}
            & \boldsymbol{u}^\star(t) = \underset{\boldsymbol{x}(t),\boldsymbol{u}(t)}{\arg \max} 
            \int_{t_0}^{t_0 + N_p} J(\boldsymbol{y}(t),\boldsymbol{u}(t)) + \boldsymbol{\dot{u}}(t)^T \boldsymbol{R} \ \boldsymbol{\dot{u}}(t)dt\\
            & \text{s.t. } \quad \text{on    } t \in [t_0,t_0 + N_p] \\
            & \boldsymbol{\dot{x}}(t) = \boldsymbol{f}(\boldsymbol{x}(t), \boldsymbol{u}(t) \ |\boldsymbol{\hat{\theta}}(t)), \quad\quad  \boldsymbol{x}(t_0) = \boldsymbol{\hat{x}}\\
            & \boldsymbol{y}(t) = \boldsymbol{h}(\boldsymbol{x}(t), \boldsymbol{u}(t)) \\
            & \boldsymbol{g}(\boldsymbol{y}(t),\boldsymbol{u}(t)) \leq 0\\
            & \lvert \boldsymbol{\dot{u}}(t) \rvert \leq \boldsymbol{\dot{u}}_{\max} 
    \end{aligned}
\end{equation}

All the symbols were previously defined, except the input movement $\boldsymbol{\dot{u}}$. $\boldsymbol{\hat{x}}$, $\boldsymbol{\hat{\theta}}$ are the current estimates of the states and parameters, and $\boldsymbol{u}_p$ the current implemented inputs. For the implementation, the system is discretized in time using, for example, orthogonal collocation on finite elements, and the input signal $u(t)$ is assumed to be piecewise constant on these elements \citep{biegler:2007}. Then, the solution $\boldsymbol{u}^\star(t)$ can be represented by a finite sequence $\boldsymbol{U}^\star = [\boldsymbol{u_0}^\star,\boldsymbol{u_1}^\star,\ldots]$. The problem is then solved repetitively at each sampling time of the system, and only the first control move in $\boldsymbol{U}^\star$ is implemented. 

\section{Case study: Subsea oil well network}
\begin{figure*}
	\centering
	\includegraphics[trim= 0.7cm 0.3cm 0.8cm 3cm, clip=true,scale =0.26]{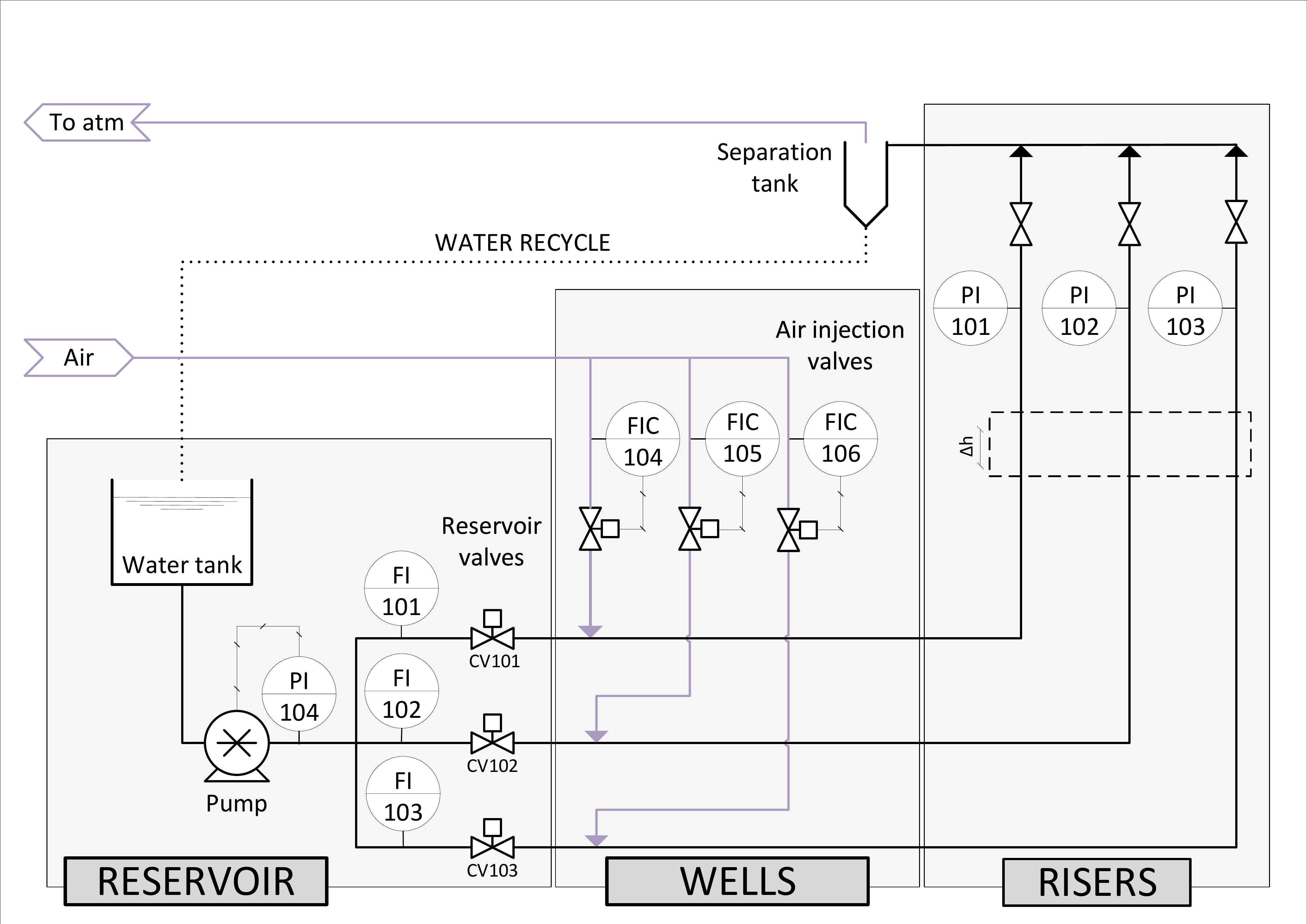}
	\caption{Experiment schematic. The system measurements $\boldsymbol{y}_p$ are the well top pressures (PI$101$, PI$102$ and PI$103$), the pump outlet pressure (PI$104$), the liquid flowrates (FI$101$, FI$102$, and FI$103$), and the gas flowrates (FI$104$, FI$105$, and FI$106$). Three PI controllers are used for controlling the gas flowrates, whose setpoints are the system inputs $\boldsymbol{u}$ computed by the optimization layer. The reservoir valve openings (CV$101$, CV$102$, and CV$103$) are the system disturbances. They change during the experiments for representing different reservoir behaviors, while the pump outlet pressure is kept constant by a PI controller.}
	\label{fig:schematic}
\end{figure*}

In subsea oil production, the goal is to extract oil and gas trapped in subsea geological structures, which is achieved by drilling several wells in these hydrocarbon reservoirs. The production of the well network is led to processing facilities on sea level by long vertical pipelines, known as risers. These facilities are responsible for separating the reservoir outflow fluids, typically gas, oil and water. 

Typically, the reservoir pressure drives the fluids from below the seafloor to the top facilities. If this pressure is not large enough, artificial lifting methods, such as electrical submersible pumps, subsea boosting stations, and gas lifting, need to be applied. The latter is commonly used since it has a robust design and relatively low-cost \citep{ben:2017}. 

\begin{figure}[ht]
	\centering
	\includegraphics[width=0.30\textwidth]{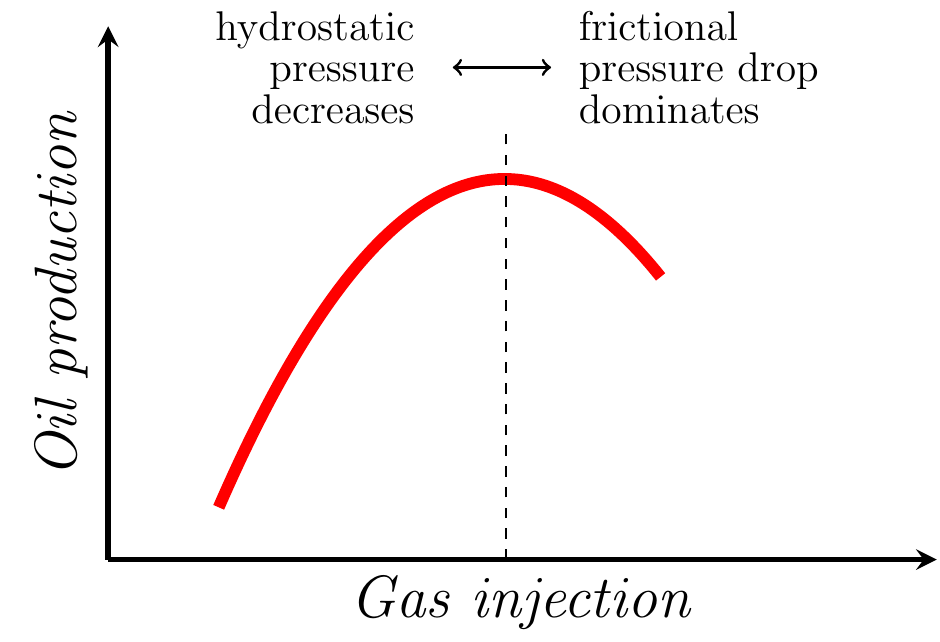}
	\caption{Gas lift effect on well's oil production. The dashed line indicates the turning point, in which the frictional pressure drop effect starts to dominate the hydrostatic pressure.}
	\label{fig:gasLiftEffect}
\end{figure}

In gas lifted systems, the excess gas that is produced is compressed and injected back in the well \citep{hernandez:2016}. As a result, fluid bulk density is reduced, decreasing the hydrostatic bulk pressure on the reservoir and increasing the production. Since the reservoirs can be located several kilometers below sea level, such a decrease has significant positive effect on the system productivity. However, if the gas injection flowrate becomes too large, the effect of frictional pressure drop in the pipelines dominates, decreasing the gain of injecting more gas to the system. The gas lift effect is illustrated by Figure~\ref{fig:gasLiftEffect}.

On a daily basis, operators and engineers are responsible for deciding the gas lift injection flowrates such that the system operates as efficiently as possible, maximizing the revenue \citep{foss:knudsen:grimstad:2018}. This decision process can be automated and improved by the usage of production optimization approaches, such as \SSRTO, \ROPA, and \DRTO.

\subsection{Experimental rig}

To study how the different approaches perform in a subsea oil well network, a small-scale experimental rig was designed. For simplification purposes, the setup uses water and air as working fluids instead of oil and gas. However, the choice of the fluids does not influence the gas lift phenomenon, which can still be seen in the lab rig. A simplified flowsheet of the rig is shown in Figure~\ref{fig:schematic}. The system is divided into three sections: \medskip

\noindent\textbf{Reservoir}: this section is composed of a \SI{200}{\liter} stainless steel tank, a centrifugal pump, and three control valves (CV$101$, CV$102$, and CV$103$). We can freely manipulate the openings of these valves, which are chosen for representing different reservoir behaviors. With this setup, the reservoir produces only liquid and its outflow ranges from \SIrange{2}{15}{\liter\per\minute}. Flow meters (FI$101$, FI$102$, and FI$103$) are located before the reservoir valves. A controller regulates the pump rotation such that the its outlet pressure (PI$104$) remains at a given setpoint, which is kept constant at \SI{0.3}{\bar g} in the experiments; \medskip 

\noindent\textbf{Wells}: three parallel flexible hoses with \SI{2}{\centi\metre} inner diameters and length of \SI{1.5}{\metre} represent the wells. Approximately \SI{10}{\centi\metre} after the reservoir valves, air is injected by three air flow controllers (FIC$104$, FIC$105$, and FIC$106$) within the range of \SIrange{1}{5}{\sliter\per\minute}; \medskip

\noindent\textbf{Risers}: this section is composed of three vertical pipelines, orthogonal to the well section, with \SI{2}{\centi\metre} inner diameters and \SI{2.2}{\metre} high. We measure the pressures on top of the risers (PI$101$, PI$102$, and PI$103$). After the sensors, we have three manual valves which are kept open during the experiments. The air is vented out to the atmosphere. For environmental purposes, the liquid is recirculated to the reservoir water tank. 

\section{Experimental Setup}
The optimal operation point of the system is achieved by maximizing the ``oil'' revenue, while accounting for gas availability constraints and bounds on the gas lift flowrates. The economic objective function $J$ is chosen as:
\begin{equation}\label{eq:economiOF}
    J = 20 \ Q_{l,1} + 10 \ Q_{l,2} + 30 \ Q_{l,3}
\end{equation}

\noindent where, $Q_{l}$ are the liquid flowrates of wells 1, 2, and 3. For illustration purposes, we assumed that the wells have different valued hydrocarbons, which are reflected by different weights in $J$. The constraint set $g$ in Equation~\eqref{eq:ssopt} and \eqref{eq:dynopt} is composed by the gas lift injection $Q_{g}$ lower and upper bounds ($Q_{g,min} =$ \SI{1}{\sliter\per\minute}  and $Q_{g,max} =$ \SI{5}{\sliter\per\minute}) and the gas availability constraint ($Q_{g,1} + Q_{g,2} + Q_{g,3} \leq$ \SI{7.5}{\sliter\per\minute}). 

\begin{figure}[ht!]
    \centering
    \includegraphics[trim= 4.3cm 15.6cm 4.7cm 9cm, clip=true, width=0.48\textwidth]{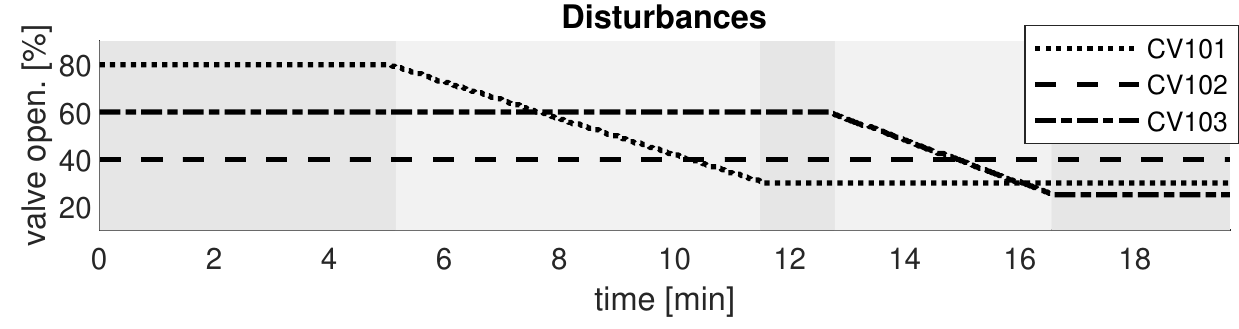}
    \caption{Disturbance profile. Changes in CV$101$ and CV$103$ openings emulate a declining oil production in wells 1 and 3.}
    \label{fig:disturbances}
\end{figure}

To study \ROPA's performance in comparison with the other approaches, we run experiments of \SI{20}{\minute} and change the opening of the valves CV$101$, CV$102$ and CV$103$ according to Figure~\ref{fig:disturbances}, while keeping the pump outlet pressure $P_{pump}$ constant. Since the holdup inside the pipes is small, the system response to changes in the gas flowrates is fast ($<$\SI{1}{\minute}). The reservoir valves are used to emulate slow time-scale and persistent reservoir disturbances. This is similar to what happens in practice with a significant time scale separation between the wells and reservoir \citep{foss:knudsen:grimstad:2018}. This disturbance scenario emulates the well's depletion (i.e. declining oil production over time), where larger valve openings indicate larger reservoir outflows. 

The initial value of the inputs is specified as $Q_{g,1} = Q_{g,2} = Q_{g,3} =$ \SI{2.5}{\sliter\per\minute}. During the experiment, we want to find the best gas lift distribution among the three wells. For carrying out this task, we implement the three production optimization approaches and a naive strategy, in which the inputs are fixed as \SI{2.5}{\sliter\per\minute} during the entire experiment. The latter is used as a baseline for the performance comparison. The procedures for implementing \ROPA, \SSRTO and \DRTO are presented in Algorithms~\ref{alg:ROPA}, \ref{alg:SSRTO}, and \ref{alg:DRTO}, respectively. The codes used in the experimental rig are available in our Github page\footnote{\url{https://github.com/Process-Optimization-and-Control/ProductionOptRig}}. In Appendix~\ref{sec:appendixTable}, we show the tuning parameters for the three approaches.
\begin{algorithm}[!ht]
\caption{\ROPA}
\begin{algorithmic}[1]
    \STATE Get plant measurements $\boldsymbol{y}_{p,k}$ and inputs $\boldsymbol{u}_{p,k}$, where $k$ indicates the current time instant\medskip
    \STATEx \textbf{Dynamic model adaptation (\EKF)}
        \State{Define an extended state with $\hat{\boldsymbol{x}}^e_k = [\hat{\boldsymbol{x}}_k,\ \hat{\boldsymbol{\theta}}_{k}]^T$}\label{alg:dynMA_1}
        \State{Use random walk for parameter evolution characterization: $\boldsymbol{\theta}_{k + 1} = \boldsymbol{\theta}_{k} +  \boldsymbol{\upsilon}_k, \quad \boldsymbol{\upsilon} \sim \mathcal{N}(\textbf{0},Q_\theta) $}\label{alg:rw}
        \State{Obtain the affine version of the dynamic model $\boldsymbol{f}(\cdot, \cdot|\cdot)$ using the sensitivity equations and extend it with the parameter evolution model from Step~\ref{alg:rw}}
        \State{Update extended state estimate covariance matrix and the estimator gain}
        \State{Compute the extended state estimate based on the current prediction error}
        \State{Obtain parameter estimates $\hat{\boldsymbol{\theta}}_{k+1}$ based on current measurements}\label{alg:dynMA_end}\medskip
    \STATEx \textbf{Steady-state economic optimization}
        \State{Update steady-state model $0 = \boldsymbol{f}(\cdot,\cdot|\boldsymbol{\hat{\theta}_k})$}
        \State{Compute $\boldsymbol{u}^\star_{k+1}$ using Equation~\eqref{eq:ssopt}}
        \State{Apply input filter $\boldsymbol{u}_{k+1} = \boldsymbol{u}_{p,k} +  K_u(\boldsymbol{u}^\star_{k+1} - \boldsymbol{u}_{p,k})$}\label{alg:ropaFilter}
        \State{Implement $\boldsymbol{u}_{k+1}$}
\end{algorithmic}
\label{alg:ROPA}
\end{algorithm}
\begin{algorithm}[!ht]
\caption{\SSRTO}
\begin{algorithmic}[1]
    \STATE Get plant measurements $\boldsymbol{y}_{p,k}$ and inputs $\boldsymbol{u}_{p,k}$\medskip
    \State Steady-state detection of $\boldsymbol{y}_{p,k}$\medskip
    \Statex{If Steady-state is $True$}\medskip
        \STATEx \textbf{$\hspace{0.6cm}$Steady-state model adaptation}
            \State{$\hspace{0.6cm}$Compute parameter estimates $\hat{\boldsymbol{\theta}}_{k+1}$ using Equation~\eqref{eq:ssmpa}}\medskip
        \STATEx \textbf{$\hspace{0.6cm}$Steady-state economic optimization}
            \State{Update steady-state model $0 = \boldsymbol{f}(\cdot,\cdot|\boldsymbol{\hat{\theta}_k})$}
            \State{$\hspace{0.6cm}$Compute $\boldsymbol{u}^\star_{k+1}$ using Equation~\eqref{eq:ssopt}}
            \State{$\hspace{0.6cm}$Apply input filter $\boldsymbol{u}_{k+1} = \boldsymbol{u}_{p,k} +  K_u(\boldsymbol{u}^\star_{k+1} - \boldsymbol{u}_{p,k})$}\label{alg:ssrtoFilter}
            \State{$\hspace{0.6cm}$Implement $\boldsymbol{u}_{k+1}$}\medskip
    \STATEx {Else}
        \State{$\hspace{0.6cm}$ Do nothing}
\end{algorithmic}
\label{alg:SSRTO}
\end{algorithm}
\begin{algorithm}[!ht]
\caption{\DRTO}
\begin{algorithmic}[1]
    \STATE Get plant measurements $\boldsymbol{y}_{p,k}$ and inputs $\boldsymbol{u}_k$\medskip
    \STATEx \textbf{Dynamic model adaptation (\EKF)}
        \State{Implement Step~\ref{alg:dynMA_1} to \ref{alg:dynMA_end} from Algorithm~\ref{alg:ROPA}}\medskip
    \STATEx \textbf{Dynamic economic optimization}
        \State{Update dynamic model $\boldsymbol{f}(\cdot, \cdot|\boldsymbol{\hat{\theta}_k})$}
        \State{Compute $\boldsymbol{U}_{k+1}^\star$ using Equation~\eqref{eq:dynopt}}
        \State{Extract $\boldsymbol{u}_{k+1}$ from $\boldsymbol{U}_{k+1}^\star$}
        \State{Implement $\boldsymbol{u}_{k+1}$}
\end{algorithmic}
\label{alg:DRTO}
\end{algorithm}

Note that we use first order input filters in the implementation \ROPA (Algorithm~\ref{alg:ROPA}/Step~\ref{alg:ropaFilter}) and \SSRTO (Algorithm~\ref{alg:SSRTO}/Step~\ref{alg:ssrtoFilter}) as well as input regularization in the objective function of the dynamic economic optimization (Equation~\ref{eq:dynopt}). Despite affecting the economic performance, penalizing large input variations is necessary for practical applications. For example, since the wells in our problem are identical, small deviations in their models, which can result from slightly different updates in their parameters, may lead to significant changes in the optimal gas lift flowrate distribution. Potentially, these values can go from the minimum to the maximum limits in one iteration. This type of decision can easily raise credibility issues on the plant engineers and operators concerning the production optimization approaches' performance. Thus, they need to be avoided and input filters/regulation are an interesting alternative. 

\section{ROPA Implementation}\label{sec:impGuide}

Before showing the main results regarding the comparison of the three approaches, we present some recommendations for \ROPA's practical implementations\footnote{In Appendix~\ref{sec:appendix}, we also briefly discuss the implementations of \SSRTO and \DRTO}. They are mostly based on lessons learned during the implementation on the experimental rig and can be used as guidelines for other systems.   They are focused on three aspects that should be defined before the online deployment: \textit{Process modeling},  \textit{Mo-del parametrization}, and \textit{\ROPA execution period tuning}. 

\subsection{Process modeling}
\subsubsection{General considerations}
The first step for implementing \ROPA is obtaining the dynamic and steady-state models. Their complexity should be carefully taken into account. The models should reflect the main economic trade-offs and accurately predict the system operating constraints. The more precise the model, the better the production optimization performance \citep{camara:quelhas:pinto:2016}. However, if the model is too complex, the model adaptation task becomes more challenging and little value is then added over a simplified version \citep{darby:nikolaou:jones:nicholson:2011}. Additionally, model maintenance also becomes harder.

\subsubsection{Experimental rig modeling}
The goal of the production optimization approaches is to determine the optimal injection rate (i.e. the \emph{setpoints} of flow controllers FIC$104$, FIC$105$, and FIC$106$) such that $J$ is maximized, while considering maximum gas availability constraints and gas injection bounds. Thus, when modeling the system, we should take the following into account: 
\begin{enumerate}[label=(\alph*),nolistsep,leftmargin=\parindent] %
    \item Since the system dynamics are mainly related to the reservoir, the wells' dynamics do not have to be considered in detail. Therefore, the momentum balance can be simplified, i.e. the pressure dynamic does not need to be taken into account given that its time scale is much faster than the reservoir time scale;
    \item The phenomenon of interest is gas lift. Hence, both the hydrostatic pressure and friction pressure loss need to be computed. However, since the pressure difference between the bottom and top of the riser is not significant, the gas cooking-off effect \citep{hernandez:2016} does not need to be considered. Thus, the pressure spatial variance is replaced by two lumped pressures, one at the bottom and another one at top of the riser;
    \item The main system constraint is gas availability. Since the gas flow rate is measured, the model precision with relation to this constraint is not an issue.
\end{enumerate}  \medskip

\begin{figure}[!ht]
	\centering
	\includegraphics[trim= 0.2cm 0cm 0.2cm 0cm, clip=true, width=0.45\textwidth]{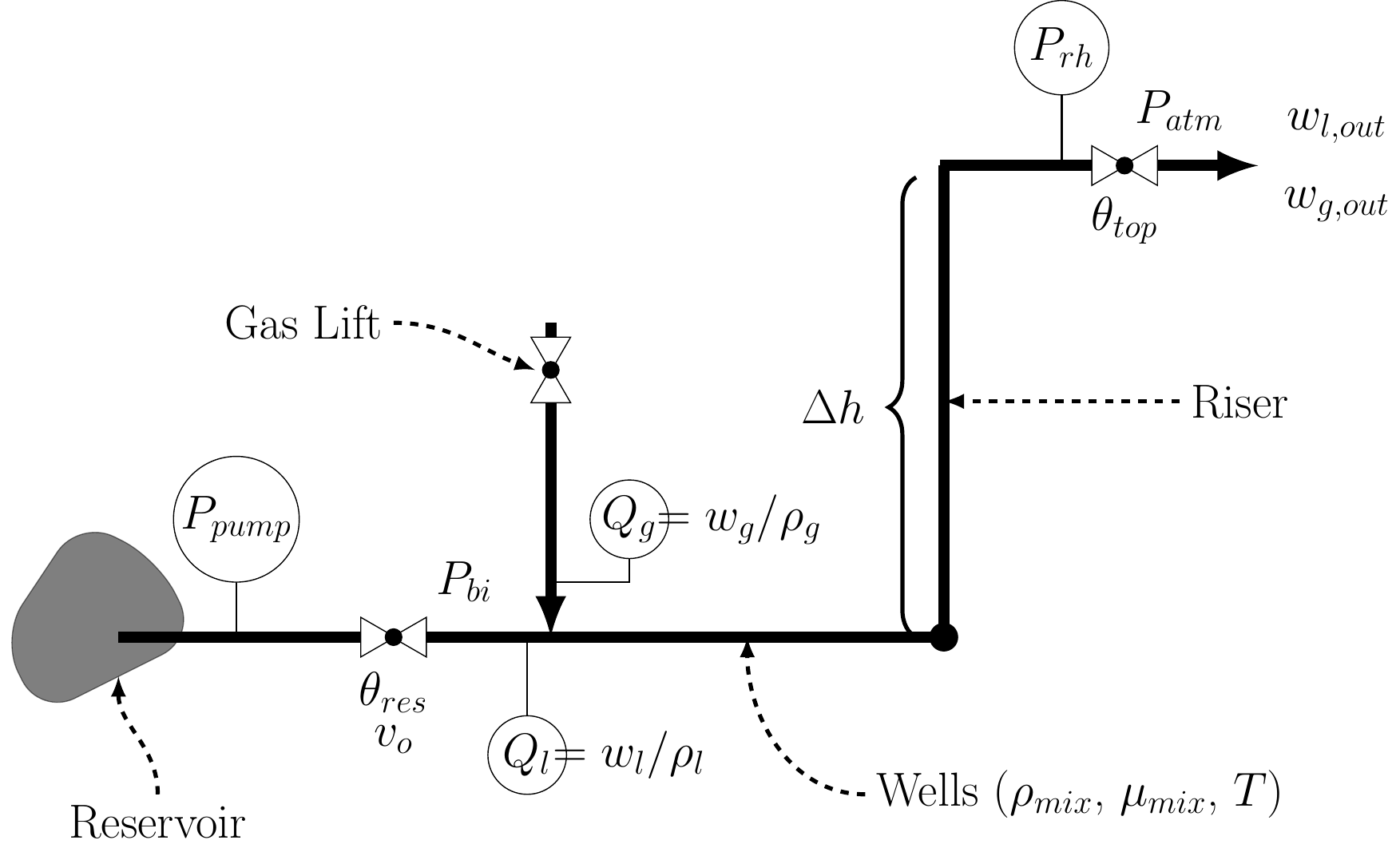}
	\caption{Diagram of a single-well model. $P_{rh}$, $P_{pump}$ and $Q_{l}$ are measured. $Q_{g}$ is controlled by the gas flowrate controller and the reservoir valve opening $v_o$ is assumed as a measured disturbance.}
	\label{fig:RigScheme}
\end{figure}

Given the considerations above, the following dynamic model is derived. It is based on \cite{krishnamoorthy:foss:skogestad:2018}. We only show the equations for one well, but the extension to three wells is straightforward. The static model is obtained by setting the time derivatives to zero. A model diagram is shown in Figure~\ref{fig:RigScheme}. 

\textbf{Differential equations.} The liquid and gas mass balances are represented by:
\begin{subequations}\label{eq:massBaldyn}
     \begin{align}
         \dot{m}_{g} & = w_g - w_{g,out} \\
         \dot{m}_{l} & = w_l - w_{l,out}
     \end{align}
\end{subequations}

\noindent where, $m_l$ and $m_g$ are the liquid and gas mass holdups inside the wells and riser. The dot symbol represents the time derivative. $w_{g}$ is the gas injection mass flowrate and $w_{l}$ is the liquid flowrate coming from the reservoir. $w_{g,out}$ and $w_{l,out}$ are the outlet production rate of gas and liquid. 

\textbf{Algebraic equations:} 
The reservoir outflow is obtained by the following relationship:
\begin{equation}\label{eq:liqFlow}
     w_l = v_o\theta_{res}\sqrt{\rho_l(P_{pump} - P_{bi})} 
\end{equation}

\noindent where, $\rho_{l}$ is the liquid density, $\theta_{res}$ is the reservoir valve ﬂow coefﬁcient, and $v_o$ the valve opening. The pump outlet pressure $P_{pump}$ is measured and the pressure before the injection point $P_{bi}$ is computed taking into account the hydrostatic pressure and the pressure drop due to friction (as a simplification, we use the Darcy-Weisbach expression for laminar flow in cylindrical pipes). Thus, $P_{bi}$ becomes:
\begin{equation}\label{eq:pressDrop}
     P_{bi} = P_{rh} + \rho_{mix}g\Delta h + \frac{128\mu_{mix}(w_{g} + w_{l})L}{\pi \rho_{mix} D^4}
\end{equation}

\noindent where, $P_{rh}$ is the pressure at the riser head, which is measured. $\Delta h$, $L$ and $D$ are the height, length (well + riser), and diameter of the pipes. $g$ is the gravitational acceleration. $\mu_{mix}$ is the mixture (liquid + gas) viscosity. In the experimental setup, the mixture viscosity is approximated as the liquid viscosity. The mixture (liquid + gas) density $\rho_{mix}$ is obtained by:
\begin{equation}\label{eq:denstotal}
     \rho_{mix} = \dfrac{{m}_{total}}{{V}_{total}}  = \dfrac{m_{g} + m_{l}}{{V}_{total}} 
\end{equation}

\noindent which is complemented by an equation indicating that the summation of the gas $V_{g}$ and liquid $V_{l}$ volumetric holdups is equal to the total system volume:
\begin{equation}\label{eq:volumeConst}
     V_{total} = V_g + V_l = \dfrac{{m}_{l}}{\rho_{l}} + \dfrac{{m}_{g}}{\rho_{g}}
\end{equation}

The liquid density $\rho_{l}$ is assumed constant, whereas the gas density $\rho_{g}$ is  computed using the ideal gas law:
\begin{equation}\label{eq:densGas}
     \rho_{g} = \dfrac{P_{bi}M_{g}}{R T}  
\end{equation}

\noindent where, $M_{g}$ is the air molecular weight, $R$ the gas universal constant, and $T$ the room temperature. The total outlet flowrate is obtained by the following relationship:
\begin{equation}\label{eq:flow}
     w_{total} =  w_{g,out} + w_{l,out} = \theta_{top}\sqrt{\rho_{mix}(P_{rh} - P_{atm})}
\end{equation}

\noindent where, $P_{atm}$ is the atmospheric pressure, and $\theta_{top}$ is the top valve ﬂow coefﬁcient. We make an additional assumption that the proportion between liquid and total outlet flowrate is the same as the liquid fraction in the mixture $\alpha_l$, i.e.:
\begin{equation}\label{eq:flowAprox}
     \alpha_l = \dfrac{m_{l}}{m_{total}} = \dfrac{w_{l,out}}{w_{total}} 
\end{equation}

\begin{figure*}
	\centering
	\includegraphics[trim= 0.5cm 3.8cm 0.1cm 3.5cm, clip=true,width=0.85\textwidth]{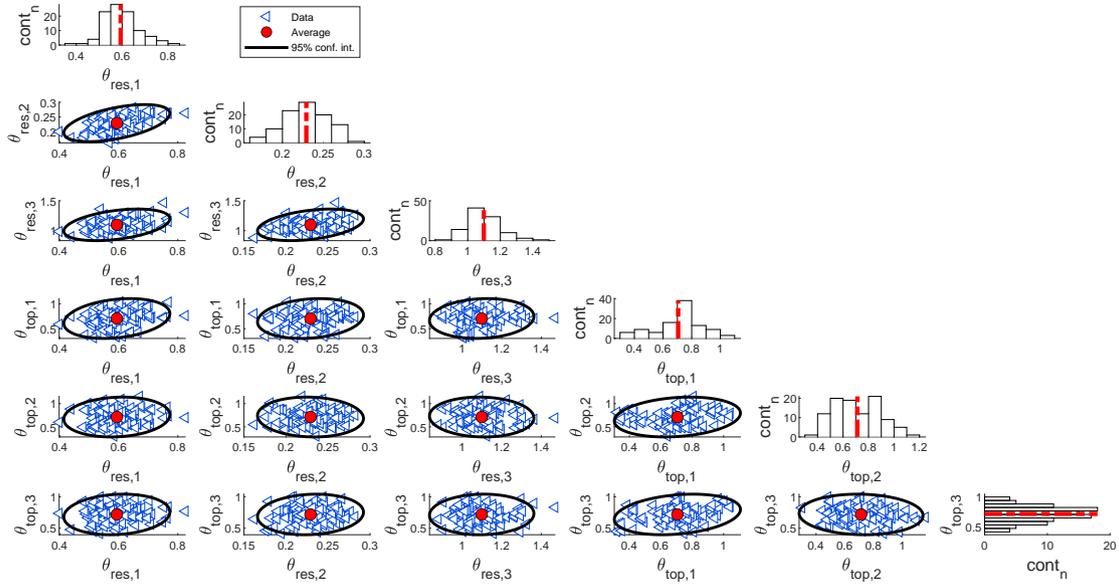}
	\caption{100 independent steady-state model adaptation runs using historical data. The histograms of the individual parameters are plotted, with a red line indicating the mean. Also, we show the 2-dimensional distribution, analyzing two parameters at a time. In these plots we also the average (red dot) and 95\% confidence interval (black line).}
	\label{fig:parameterEstimation}
\end{figure*}

\subsection{Model parametrization}
\subsubsection{General considerations}
The second step is to select which model parameters are adapted online\footnote{Note that we assume a good instrumentation design and maintenance, which can provide measurements with low noise levels and unbiased}. Despite being commonly overlooked, a poor selection can result in an ill-conditioned model adaptation step and have a significant negative impact on the production optimization economic performance \citep{quelhas:jesus:pinto:2013}. A good starting point is to select a subset of model parameters that can be adjusted such that the most significant process disturbances are represented in the model. Moreover, if the parameters change slowly (e.g. representing equipment degradation), it may not be appropriate to update them at the production optimization time scale. 

Several strategies for avoiding model adaptation problems that are ill-conditioned are available in the literature. They are typically referred to as identifiability analysis. There are different methods for testing identifiability of steady-state models (e.g. \cite{walter:pronzato:1997} and \cite{yang:wang:shao:biegler:2018}) and of dynamic models. \cite{miao:xia:perelson:wu:2011} provide a review of identifiability analysis of nonlinear \ODE models. 

Since \ROPA relies on dynamic model adaptation, methods of the second type should be used for a rigorous analysis. However, the application of these methods is challenging for larger models \citep{miao:xia:perelson:wu:2011}. In the authors' experience, we verified that guaranteeing that the steady-state model adaptation is well conditioned already suffices for most of the practical implementations when the goal is to operate around a steady-state point.

\subsubsection{Experimental rig model parametrization}
By extending Equations~\eqref{eq:massBaldyn}-\eqref{eq:flowAprox} to three wells, the following parameter set is obtained:
\begin{equation}\label{eq:modelParameters}
    \boldsymbol{\theta} = [\theta_{res,1},\theta_{res,2},\theta_{res,3},\theta_{top,1},\theta_{top,2},\theta_{top,3}]^T.
\end{equation}

\noindent Here, the notation is selected such that $\theta_{res,1}$ is the valve constant for reservoir valve at well $1$, and so on. Note that $\boldsymbol{\theta}$ could also include variables such as $\mu_{mix}$ and $\alpha_l$ from Equations~\eqref{eq:pressDrop} and \eqref{eq:flowAprox}, which would exclude the need of some model simplifying assumption, namely: mixture viscosity equals to the liquid viscosity, and the liquid ratio of the outlet flowrate is equal to the liquid fraction. Instead of making this choice based on a subjective basis, we carried out a test for making sure that the steady-state model adaptation problem was well-conditioned. Based on historical \SS data, we estimated the parameters $100$ times using the \SS model adaptation (Equation~\eqref{eq:ssmpa}). 

The distribution of these estimates of $\boldsymbol{\theta}$ is shown in Figure~\ref{fig:parameterEstimation}. We see that the computed confidence regions of the estimated parameters are bounded and small compared to their magnitudes. Also, none of the estimates lies on the constraints (which are not indicated in the plots). When $\alpha_l$ was included in $\boldsymbol{\theta}$ and we ran the same test (see Appendix~\ref{sec:poorParameterSet}), a considerable part of the estimates of $\alpha_l$ was at the constraints. Although these bounds force the parameters estimates to adequate physical ranges, such a pattern would have been a clear indication of an ill-conditioned model adaptation problem and/or improper definition of the bounds. As a consequence, the adapted parameter values would lose their physical and statistical meaning and the model updating step may become useless. 

Another characteristic of ill-conditioned model adaptation problems is parameter estimates with high correlations (i.e. main axes of the confidence interval ellipsoid in Figure~\ref{fig:parameterEstimation} are not perpendicular to the plot axes). A correlation pattern can be noted among the reservoir valve parameters ($\theta_{res,1}$, $\theta_{res,2}$, and $\theta_{res,3}$), which is mainly caused by the large influence of the pump outlet pressure in the estimated value. Since the correlation is not significant (the inclination angle is relatively small), the parameters can be individually estimated. Usually, this is the desired case; however, it may be advantageous to pose an ill-determined estimation problem for flexibility of the estimation problem (cf. Chapter 7 in \citep{bard:1974}).

\begin{remark}
If this analysis is carried out, the tuning step of the \EKF becomes much simpler, since a good initial guess for the estimate covariance matrices (both the model states and parameters) is available.
\end{remark}

\subsection{ROPA execution period tuning}\label{sec:ropatuning}
\subsubsection{General considerations}
The final step for \ROPA implementation is determining its execution period $\Delta t_{ROPA}$. For this decision, we should consider the system response time. As in any \RTO implementation, we assume that there is a control layer (either advanced or PIs controllers) implementing the set points determined by \ROPA. Therefore, we need to take their dynamic effects into account when determining $\Delta t_{ROPA}$. 

If $\Delta t_{ROPA}$ is excessively small, the system becomes sensitive to rapid changes in the disturbance and noise. In this case, the inputs computed by \ROPA can oscillate significantly before converging to the optimum and, in the worst-case scenario, they may not be physically realizable. On the other hand, if $\Delta t_{ROPA}$ is too large (on the order of the plant's settling time), \ROPA performance approaches the \SSRTO performance. Based on the authors' experience, $\Delta t_{ROPA}$ can be initially chosen as half of the difference between the control layer response time and the smallest plant time constant, in which only the \RTO-relevant dynamics are considered.

\subsubsection{Choosing $\Delta t_{ROPA}$ for the rig implementation}
\begin{figure}[!ht]
	\centering
	\includegraphics[trim= 4.1cm 14cm 4.7cm 8.8cm, clip=true, width=0.45\textwidth]{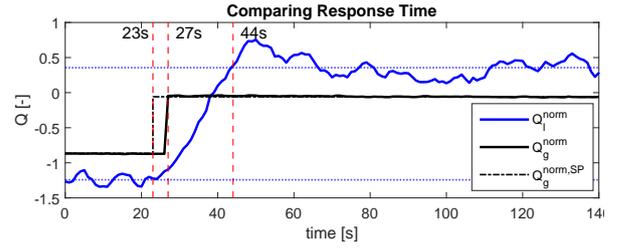}
	\caption{Step response. The initial and final mean values of $Q_{l}$ are also indicated. The values were normalized for facilitating the visualization. We step the gas injection setpoint at \SI{23}{\second}. The control takes \SI{4}{\second} to track the new setpoint, while the system settles at a new steady-state after approximately \SI{20}{\second}.}
	\label{fig:DTropa}
\end{figure}

For determining the system response time in the experimental rig, we apply a step-change in the setpoint of the gas lift flowrate controller in one of the wells and analyze the effect on the liquid flowrate. In Figure~\ref{fig:DTropa}, we show the normalized profiles of $Q_g^{SP}$, $Q_g$ and $Q_l$. For reference, we indicate the time of the setpoint step change (\SI{23}{\second}); PI controller response (\SI{27}{\second}); and system response (approx. \SI{44}{\second}). The results show that the difference between the plant and control time response is around \SI{20}{\second}. Thus, according to the rule of thumb, choosing $\Delta t_{ROPA} =$ \SI{10}{\second} is a good tuning. 

\begin{figure*}
	\begin{center}
	    \subfloat[Estimated parameter - scaled reservoir valve coefficient.]{\includegraphics[trim= 4.3cm 9cm 4.7cm 9cm, clip=true, width=0.47\textwidth]{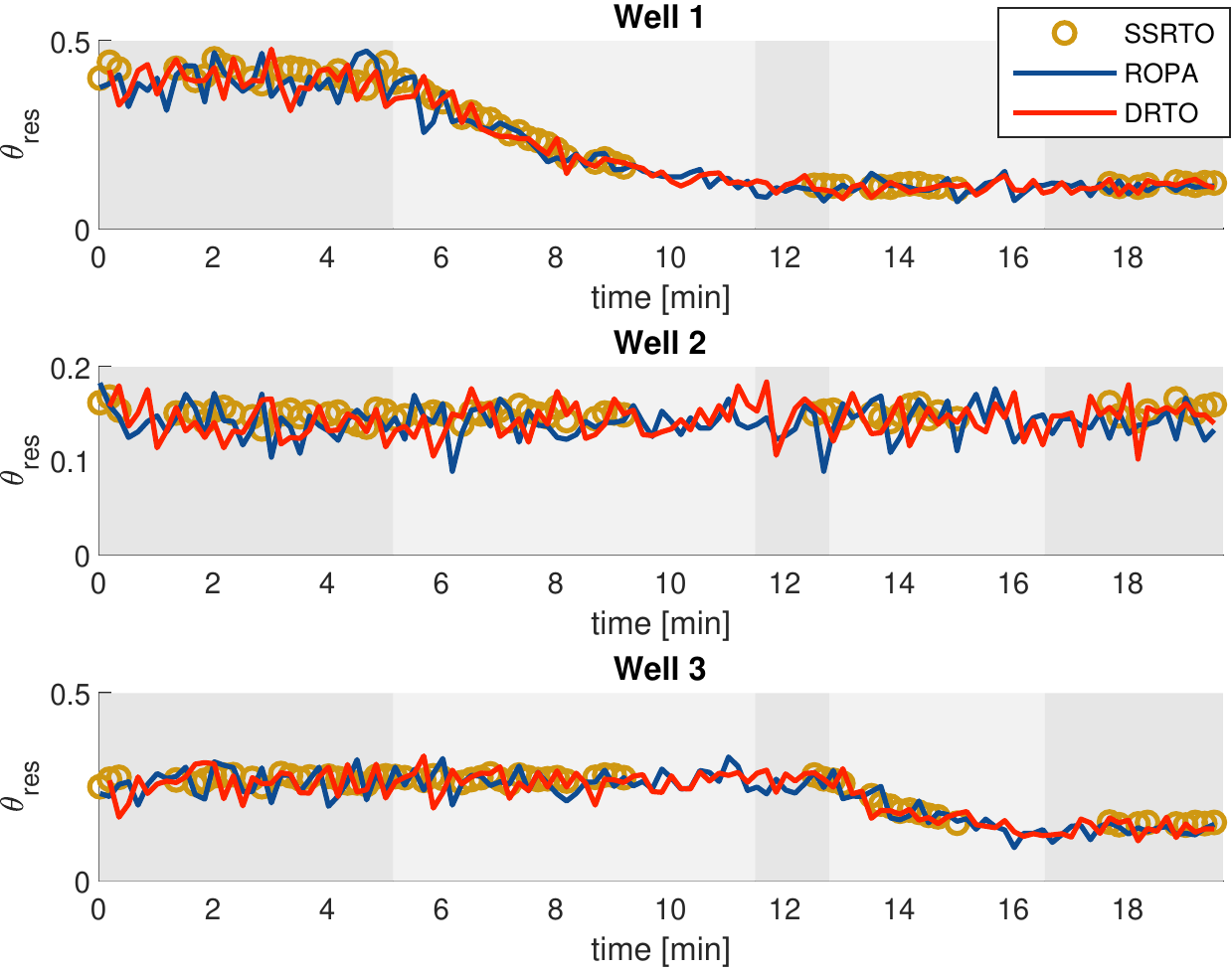}\label{fig:parameters}}
	    \subfloat[Estimated parameter - scaled top valve coefficient.]{ \includegraphics[trim= 4.3cm 9cm 4.7cm 9cm, clip=true, width=0.47\textwidth]{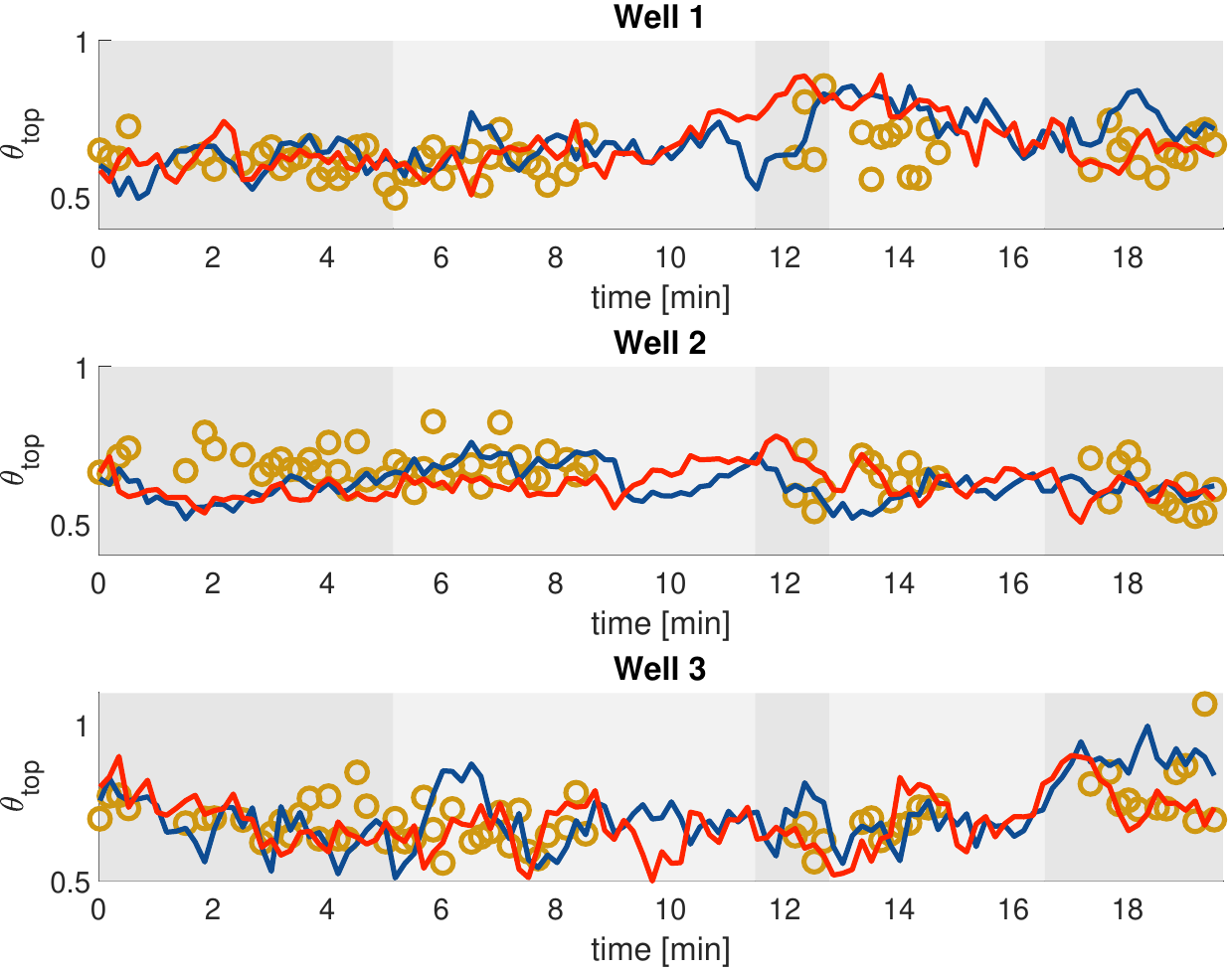}\label{fig:parameters2}}\\
	    \subfloat[Average gas lift flowrate (\SI{1}{\sliter\per\minute} $\leq Q_{g} \leq$ \SI{5}{\sliter\per\minute}). \\ Note that the scales of the three plots are different.]{\includegraphics[trim= 4.2cm 9cm 4.7cm 9cm, clip=true, width=0.47\textwidth]{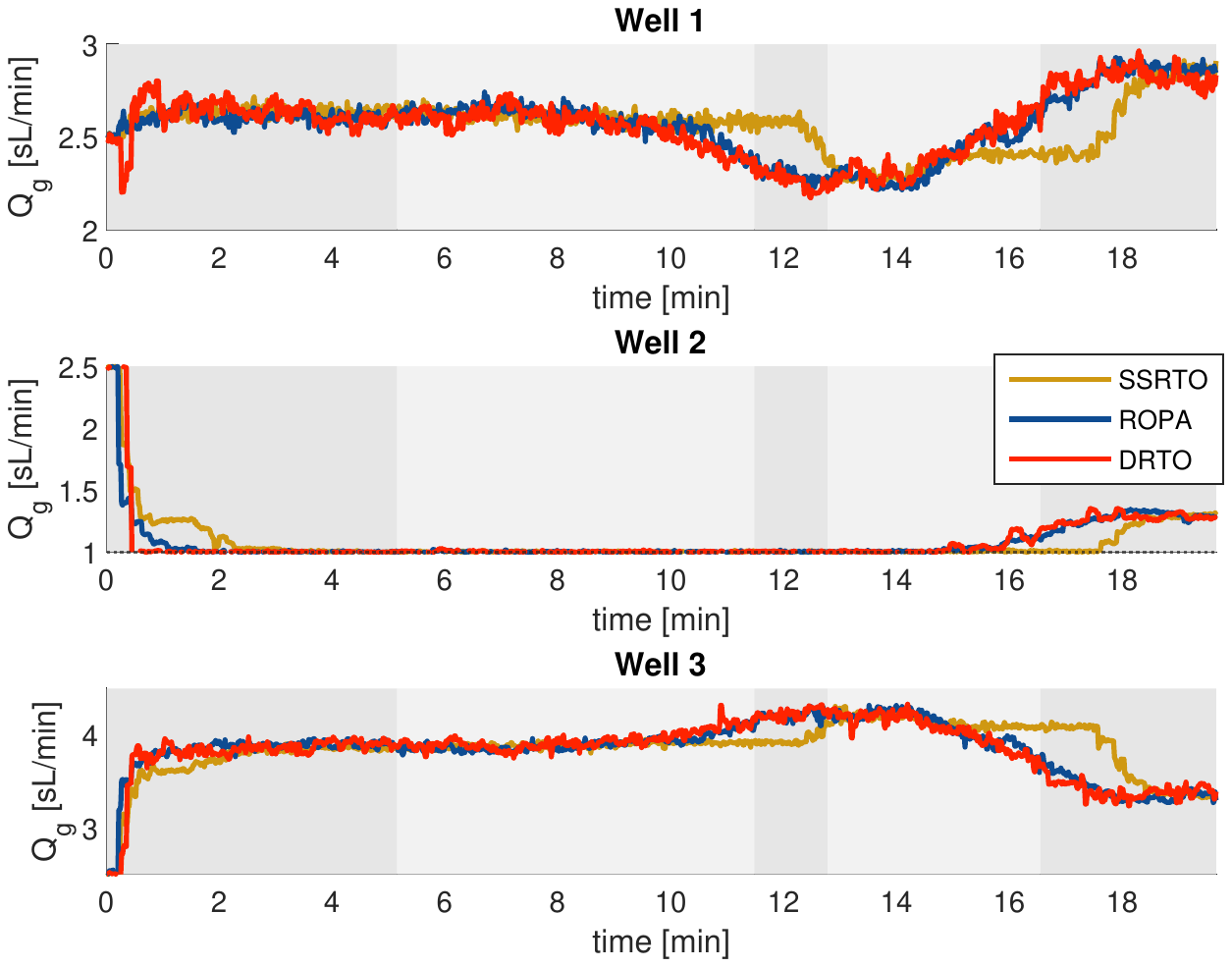}\label{fig:MV}}
	    \subfloat[Average profit. The bottom figure shows the cumulative profit difference.]{\includegraphics[trim= 4.2cm 9cm 4.7cm 9cm, clip=true, width=0.47\textwidth]{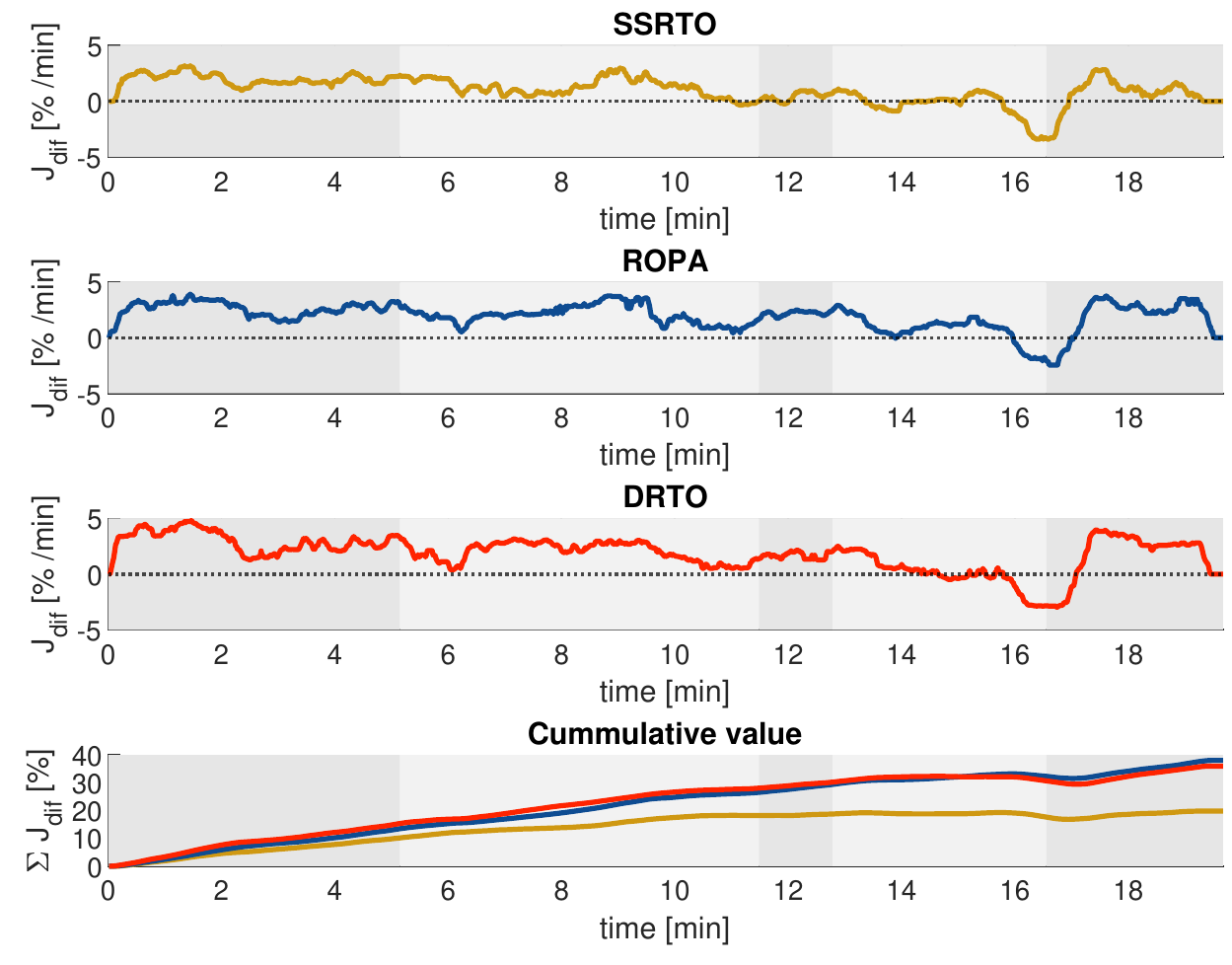}\label{fig:ProfitDif}} \\	    
	    \caption{Experimental results.}
		\label{fig:ExperimentalResults}
	\end{center}
\end{figure*}

\section{Experimental results and discussion}

The results comparing the three approaches are presented in Figure~\ref{fig:ExperimentalResults}. As shown previously, we defined $\Delta t_{ROPA}$ as \SI{10}{\second}. For comparison reasons, both \SSRTO and \DRTO are executed at the same rate. Thus, \SSRTO results can be used as baseline for performance comparison. In turn, if \ROPA and \DRTO present similar results, it is an indication that there are no major advantages in optimizing the transients in the system of interest. 
In order to mitigate the effect of noise and unmeasured disturbances in the analysis, we run two independent experiments for each approach and present the average profiles of the computed inputs and profit. On the other hand, for the estimated parameters, we show the results of a single run. The reason is that \SSRTO is triggered in different time instants during different experiments due to the \SSD procedure. Therefore, showing the average at a given time instant is not relevant. For completeness, we show the profiles of the estimated parameter of two independent \SSRTO runs in Appendix~\ref{sec:appendix}.

In Figures~\ref{fig:parameters} and \ref{fig:parameters2}, we present the values of the estimated reservoir valve and top valve coefficients. Since in \ROPA and \DRTO, we estimate the parameter every execution time (i.e. every \SI{10}{\second}), we use a continuous line. In turn, for \SSRTO, we use markers to indicate not only the estimated values but also when it was executed. The approaches show a consistent $\theta_{res}$ estimation profile with low variability of both steady-state and dynamic model adaptation steps. This is a direct consequence of the proper choice of the model parametrization, which means that the parameter values are not greatly affected by measurement noise. Moreover, the reservoir parameter profiles reflect the disturbances shown in Figure~\ref{fig:disturbances}, which is in line with the guidelines that indicate that the chosen parameters should reflect the most significant process disturbances. Note that despite the fact that both \ROPA and \DRTO use the same dynamic estimation method, the parameters were estimated in two independent runs. Thus, there are small deviations between the two profiles.

Figure~\ref{fig:MV} shows the average profile of the manipulated variable $Q_{gl}$. Similarly to the parameters results, the manipulated variables profiles are consistent among the three approaches. Moreover, the computed inputs are also consistent with the profiles inferred from engineering insight. In an exploratory experiment, we determined that lower reservoir valve openings are connected to larger input/output gains (for example, adding one unit of gas lift flowrate when $v_o =$ \SI{40}{\percent} increases the liquid flowrate more than when $v_o =$ \SI{80}{\percent}). Therefore, between \SIrange{0}{4}{\minute}, we expected the approaches to increase production in wells $1$ and $3$ due to the larger weights in the objective function. However, the fact that well $1$ has a larger input/output gain should be considered. At the end of the experiment, after \SI{17}{\minute}, all three wells have similar input/output gains. Thus, we expect the approaches to distribute the available gas more evenly among the three wells, taking only the different weights in $J$ into account.  

The profiles in Figure~\ref{fig:MV} confirm that the methods follow the expected behavior. In the beginning, well $3$ is prioritised, but its value is not kept at the maximum ($Q_{g,\max} =$ \SI{5}{\sliter\per\minute}). Since the first well has a larger input/output gain, the methods choose to balance the gas injection between wells $1$ and $3$. In turn, the gas injection in well $2$ is kept at the minimum ($Q_{g,\min} =$ \SI{1}{\sliter\per\minute}). At the end, the gas is more evenly distributed but still with $Q_{g,3} > Q_{g,1} > Q_{g,2}$, which follows the weights on $J$. 

We also note that, on average, \SSRTO is much slower to adapt the inputs in face of disturbances. The main changes in $Q_{gl}$ are made only after the disturbances stops, around \SI{12}{\minute} and \SI{18}{\minute}. On the other hand, \ROPA and \DRTO quick-ly adapt the inputs to the disturbance. Additionally, we see that \ROPA and \DRTO input profiles are mostly overlapped, which indicates that there is not much advantage in optimizing the transients on this system. The main reason is that the well time constants are much faster than the disturbances time constants. This is in line with the behavior of actual subsea oil wells, where the system dynamics are mainly determined by the reservoir dynamics \citep{foss:knudsen:grimstad:2018}.  

Figure~\ref{fig:ProfitDif} compares the profit obtained by the three approaches with the naive strategy, in which available gas lift is equally divided among the three wells (i.e. $Q_{gl,1} = Q_{gl,2} = Q_{gl,2} =$ \SI{2.5}{\sliter\per\minute}). Instead of showing absolute values, we plot the difference, in percentage, between the instantaneous profit the approach of interest and fixed input approach which is calculated as $100(J - J_{fix})/J_{fix}$. In addition, we use a \SI{60}{\second} moving average for smoothing the profiles, because the instantaneous profit measurements are noisy. 

As expected, \ROPA and \DRTO have a better performance than \SSRTO due to the higher frequency of the production optimization execution. All the approaches present better economic results than the fixed input strategy, except around \SI{17}{\minute}. In this case, an equal gas distribution yields better performance. However, since we apply input filters in \ROPA/\SSRTO and input usage regulation in \DRTO, the approaches are not able to increase the inputs to this level so rapidly. 

Next, we computed the total profit difference. We see \ROPA and \DRTO increase the obtained profit by approximately \SI{38}{\percent} and \SSRTO by \SI{20}{\percent} when compared to the fixed nominal input approach. Such improvements represent a significant advantage of the production optimization approaches. Regarding the average instantaneous profit improvement, \ROPA and \DRTO are around \SI{1.8}{\percent}, whereas \SSRTO around \SI{1}{\percent}. These values are in accordance to the profit improvement achieved in real systems by production optimization approaches \citep{foss:knudsen:grimstad:2018}. 

\begin{figure}[!ht]
	\centering
	\includegraphics[trim= 4.1cm 15.6cm 4.7cm 9.5cm, clip=true, width=0.45\textwidth]{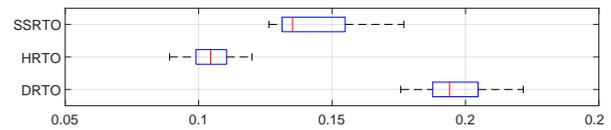}
	\caption{Distribution of the computation time of the three approaches. The average computational time values are: $\bar{\tau}_{SSRTO} =$ \SI{0.1437}{s},$\bar{\tau}_{ROPA} =$ \SI{0.1090}{s}, $\bar{\tau}_{DRTO} =$ \SI{0.2005}{s}. All computational are carried out with an Intel Core i7-8650U CPU at 2.8 GHz and 16GB RAM.}
	\label{fig:CompuTime}
\end{figure}

Finally, we compare the three approaches in terms of computational efficiency in Figure~\ref{fig:CompuTime}, where we show the computation time distribution in one experimental run. The results show that \ROPA has an average computational time approximately two times smaller than \DRTO. Even though the computational times are small here, they can steeply increase with the system size and complexity. Therefore, it can be advantageous to use \ROPA in online implementations due to its lower computational time and similar economic performance. Note that the computational time of \SSRTO is higher than \ROPA. The main reason is the parameter estimation step. While in \ROPA we use a recursive method, which is computationally cheap, in the steady-state \RTO we solve an optimization problem. However, even in this case, the maximum \SSRTO execution time is in the same order of magnitude that the minimum execution time of \DRTO, illustrating how computationally expensive is to solve a dynamic optimization problem.   

\section{Conclusion}

In this paper, we show the implementation of Real-time Optimization with Persistent Parameter Adaptation (\ROPA) on a small pilot scale plant. \ROPA is an \RTO variant, in which the steady-state wait is avoided by replacing the steady-state model parameter adaptation step by a dynamic estimator.

The main contribution of this paper regards \ROPA's performance. Our paper compared \ROPA to a classical Steady-state \RTO (\SSRTO) and a Dynamic \RTO (\DRTO) in terms of parameter estimation performance, input profile, and total profit. The results show that \ROPA has an economic performance similar to \DRTO in the system of interest without the need of optimizing the plant transients. Therefore, \ROPA becomes an interesting alternative for systems that have optimal operation around a steady-state, since it optimizes the system much more frequently than \SSRTO but does not require the solution of a dynamic economic optimization like in \DRTO.  
As secondary contributions, we presented some guidelines for the practical implementation of \ROPA. Instead of emphasizing only the applied mathematics and advanced algorithms of \ROPA, we used domain knowledge about the small pilot-scale plant in order to illustrate some decision that need to be made regarding modeling, model parametrization, and \ROPA execution period. We believe that our paper moves \ROPA closer to actual implementation, increasing its potential industrial impact.

The main challenge is that \ROPA still requires a dynamic model for parameter estimation. Future research will focus on replacing \ROPA's dynamic model update step. The idea is to first obtain simplified empirical dynamic models, in which only the dominant time constants are considered (i.e. only the main dynamic effects and holdups are modeled). Then, use the steady-state prediction of these simplified models in a steady-state model adaptation step, thereby avoiding the development of the rigorous dynamic model.

\bibliographystyle{cas-model2-names}

\bibliography{refAbbreviations,refArticles,refBooks,refProceedings}

\printcredits

\appendix
\section{Extra details about SSRTO and DRTO implementation on the rig}\label{sec:appendix}
\subsection{Steady-state \RTO}
\begin{figure}[ht!]
    \begin{center}
    \subfloat[\SSD results from one experimental run]{
    \includegraphics[trim= 4.3cm 9cm 4.7cm 12.3cm, clip=true, width=0.45\textwidth]{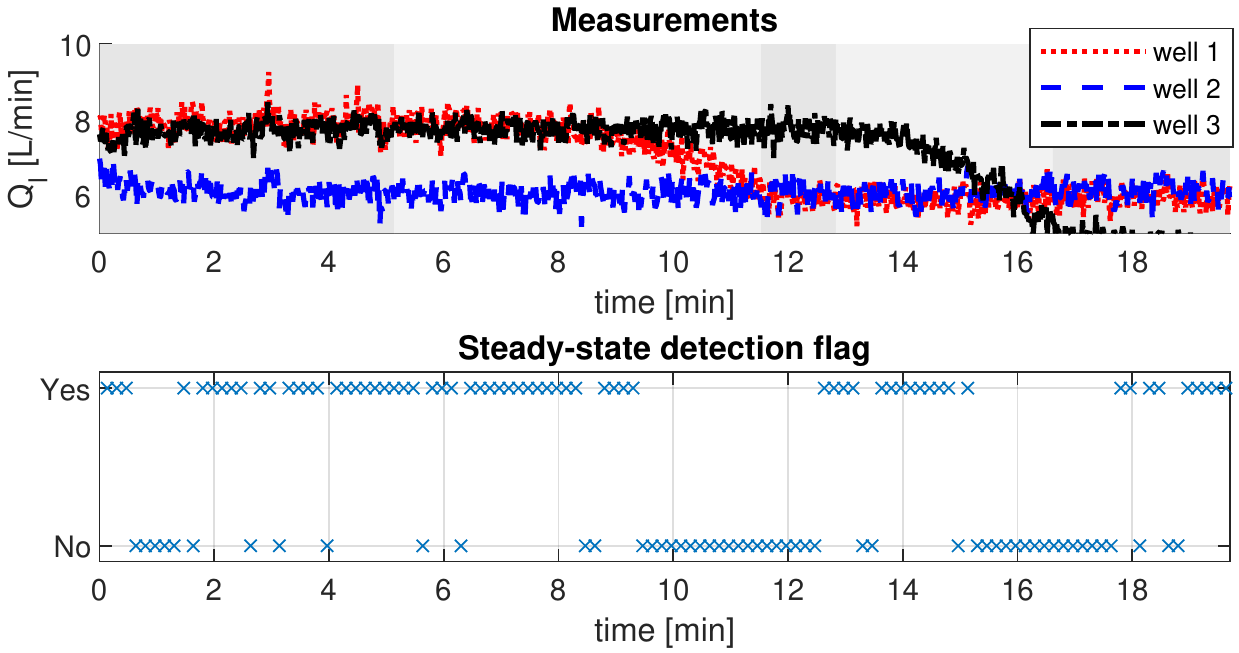}\label{fig:SSD}}\\
     \subfloat[Comparing the \SS flag of three different runs.]{\includegraphics[trim= 4.3cm 12.4cm 4.7cm 12.4cm, clip=true, width=0.45\textwidth]{figs/Results/Results_SSDet.pdf}\label{fig:SSD_2}}
    \caption{Steady-state detection procedure analysis}
    \end{center}
\end{figure}

The first step of the \SSRTO implementation is the steady-state detection (see Algorithm~\ref{alg:SSRTO}). We use the liquid flowrates $Q_{l,1}$, $Q_{l,2}$, and $Q_{l,3}$ as \SS representative measurements. For performing \SSD, we carry out a linear regression over a data window of $N_{SSD} =$ \SI{40}{\second} for each of the measurements. Next, we perform a hypothesis test (with $\alpha_{SSD} = 95\%$) on the linear models' slope parameter, in which the null hypothesis is that the slope is equal to zero. If the test fails to reject the null hypothesis, the measurement is tagged as steady-state. We consider that the system is at \SS if all three measurements pass the test. The results of one experimental run are shown in Figure~\ref{fig:SSD}. The linear regression method presented consistent results in $3$ independent runs as shown in Figure~\ref{fig:SSD_2}.

Despite not being widely used in practice \citep{cao:rhinehart:1995}, this \SSD procedure is easy to tune. The tuning parameter are the window length $N_{SSD}$ and the hypothesis test power $\alpha_{SSD}$. From a practical point of view, \SSD procedures with fewer tuning parameters are preferable, since a poor tuning choice can significantly affect the statistical foundation of the methods and, consequently, the \SSRTO results \citep{quelhas:jesus:pinto:2013}. 

\subsubsection{Parameter Estimates}\label{sec:twoSSRTOruns}
\begin{figure}[ht]
	\begin{center}
	    \subfloat[Estimated parameter - scaled reservoir valve coefficient.]{\includegraphics[trim= 4.3cm 9cm 4.7cm 9cm, clip=true, width=0.47\textwidth]{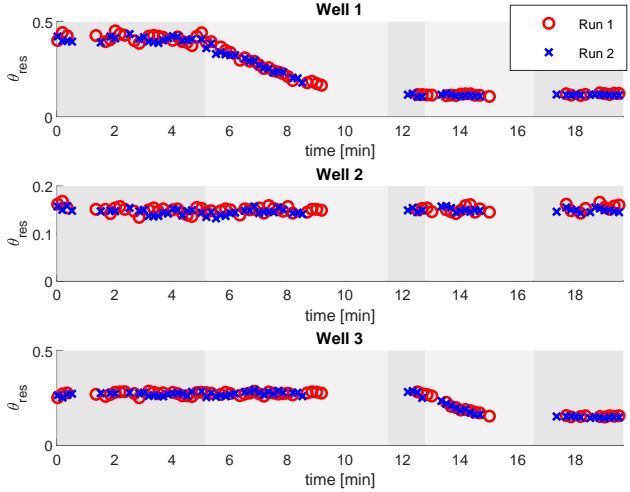}} \\
	    \subfloat[Estimated parameter - scaled top valve coefficient.]{ \includegraphics[trim= 4.3cm 9cm 4.7cm 9cm, clip=true, width=0.47\textwidth]{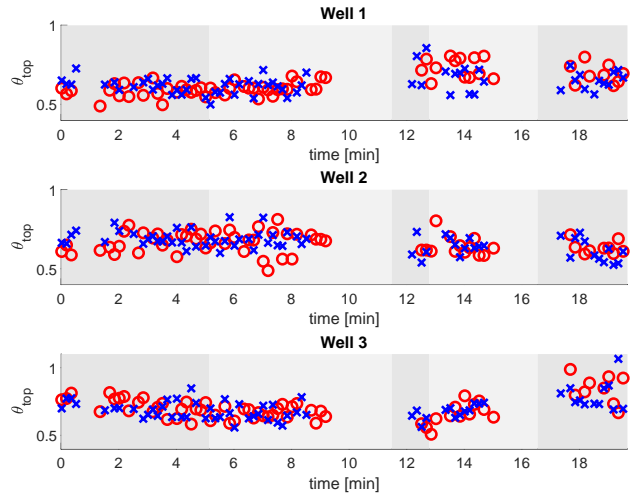}}
	    \caption{Parameter profiles for two independent \SSRTO runs. Run 2 was the one shown in Figure~\ref{fig:ExperimentalResults}}
	    \label{fig:twoSSRTOparameterRuns}
	\end{center}
\end{figure}

In Figure~\ref{fig:twoSSRTOparameterRuns}, we show the value of the profiles of the estimated parameter of two independent \SSRTO runs. We can see that the profiles are coherent and the variance of the estimates is relatively low. In the plot, we can also see the reason why we do not show the average values of the parameters in Figure~\ref{fig:ExperimentalResults}. Since the steady-state \RTO execution times are not the same in the two runs, using the average value for a given time instant is irrelevant.

\subsection{Dynamic \RTO}
Note that, in Equation~\eqref{eq:dynopt}, we have an input movement regulation term in the objective function, and a constraint on the input movement. By adding both terms, we guarantee smooth input profiles as seen in Figure~\ref{fig:MV}. However, two extra parameters $\Delta \boldsymbol{u}_{\max}$ and $\boldsymbol{R}$ need to be tuned. The former is easier to define given that it has physical meaning. Finding proper values of $\boldsymbol{R}$ is more challenging. Besides, we want to assure that the chosen tuning yields a similar input usage to the other approaches, which allows us to make a fair comparison. 

\begin{figure}[ht!]
    \centering
    \includegraphics[trim= 4.3cm 8.5cm 4.7cm 8.5cm, clip=true, width=0.38\textwidth]{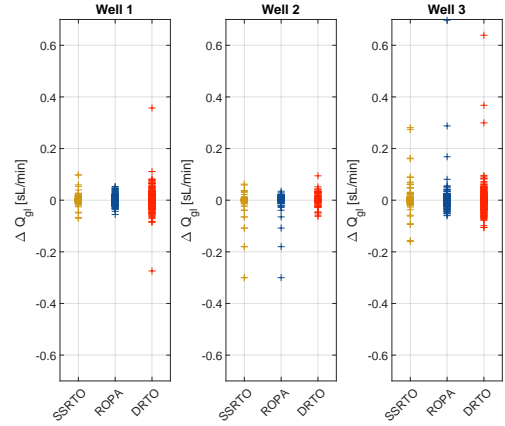}
    \caption{Comparing the input usage of the three approaches. The values indicate that both the input filter (\ROPA and \SSRTO) and the input regularization have similar performances.}
    \label{fig:inputUsage}
\end{figure}

After a trial-and-error process, the values shown in Table~\ref{tab:tuning} are used. For checking the inputs usage, we calculate the input changes $\Delta Q_{g}$ along one experimental run. We repeat the procedure for \SSRTO and \ROPA. We show the distribution of $\Delta Q_{g}$ for the three approaches in a box plot (Figure~\ref{fig:inputUsage}). We see that input usage of the \DRTO is in a similar level similar to \SSRTO and \DRTO. 

\section{Tuning parameters}\label{sec:appendixTable}
Table~\ref{tab:tuning} shows the tuning parameters used in the implementation of the three approaches. Refer to the codes available on our Github page\footnote{\url{https://github.com/Process-Optimization-and-Control/ProductionOptRig}\label{foot}} for a complete overview of how the methods are implemented. There, the methods are implemented in a high-fidelity dynamic model (digital twin) of the rig, where low level controller dynamics were also included and the noise levels were tuned according to the information obtained from the rig.
\begin{table}[!htbp]
    \caption{Tuning parameters. $I_s$ indicates an identity matrix of size $s$.}
    \resizebox{0.45\textwidth}{!}{
    \centering
    \begin{tabular}{ccc}
        \toprule
        Description & Variable & Value \\
        \midrule
        Experimental rig sensors sampling time & $T_{s}$ & \SI{1}{\second} \\\midrule
        \multicolumn{3}{c}{\ROPA} \\
        \midrule
        Execution periods & $\Delta t_{ROPA}$ & \SI{10}{\second} \\
        \ROPA Input filter gain & $K_{u}$ & \SI{0.4}{} \\
        \EKF parameters & \multicolumn{2}{c}{see Codes in Github} \\\midrule
        \multicolumn{3}{c}{\SSRTO} \\
        \midrule
        \SSD execution period & $\Delta t_{ROPA}$ & \SI{10}{\second} \\
        \SSD hypothesis test power & $\alpha_{SSD}$ & \SI{0.9}{} \\
        \SSD measurement window length & $N_{SSD}$ & \SI{40}{\second} \\
        \SSRTO Input filter gain & $K_{u}$ & \SI{0.4}{} \\
        Model adaptation weighting matrix & $V$ & $I_6$ \\
        \midrule
        \multicolumn{3}{c}{\DRTO} \\
        \midrule
        \DRTO sampling time & $T_p$ & \SI{10}{\second}  \\
        Prediction horizon & $N_p$ & \SI{6}{}   \\
        Input movement regulation & $R$ & $0.01$ \\
        Max input change & $\Delta u_{\max}$ & \SI{2}{\sliter\per\minute} \\
        \bottomrule
    \end{tabular}}
    \label{tab:tuning}
\end{table}

\newpage

\section{Model Parameter Estimation When Liquid Fraction is Included} \label{sec:poorParameterSet}
\begin{figure*}
	\centering
	\includegraphics[trim= 0.5cm 3.8cm 0.1cm 3.5cm, clip=true,width=0.95\textwidth]{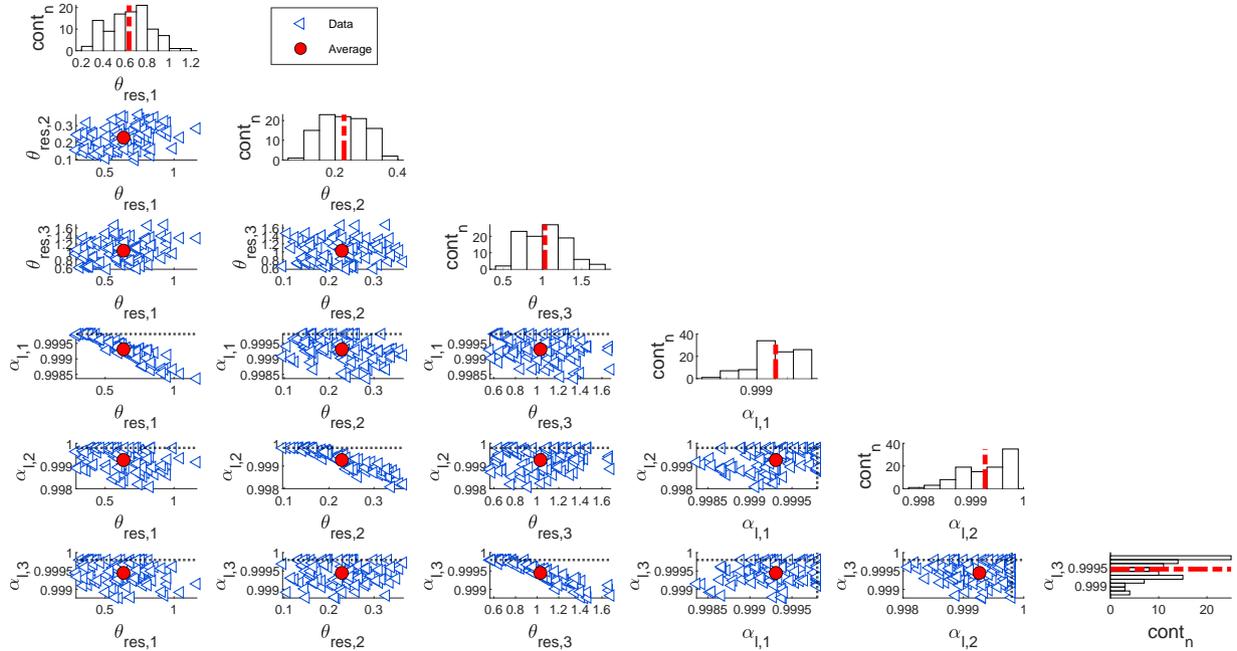}
	\caption{100 independent steady-state model adaptation runs using historical data. The histograms of the individual parameters are plotted, with a red line indicating the mean. Also, we show the 2-dimensional distribution, analyzing two parameters at a time. In these plots we also the average (red dot).}\label{fig:poorParameterSet}
\end{figure*}

In Figure~\ref{fig:poorParameterSet}, we illustrate the effect of choosing a different estimable parameter set $\boldsymbol{\theta}$. In this case, instead of estimating the top pressure valve coefficient, we included the liquid fraction in the pipelines, $\alpha_l$, for each one of the wells in the set $\boldsymbol{\theta}$: 
\begin{equation}\label{eq:modelParameters}
    \boldsymbol{\theta} = [\theta_{res,1},\theta_{res,2},\theta_{res,3},\alpha_{l,1},\alpha_{l,2},\alpha_{l,3}]^T.
\end{equation}

Consequently, we did not need the model assumption related to the liquid ratio (i.e. the outlet flowrate liquid fraction is equal to $\alpha_l$. Then, we ran the same test as in Figure~\ref{fig:parameterEstimation}. The top valve flow coefficients were set to a nominal value during the test.

First, there is a clear correlation pattern between $\theta_{res}$ and $\alpha_{l}$ of the same well. Secondly, some of the estimates $\alpha_l$ lie on the constraints. One could potentially set new values for the upper bounds; however, this test shows that the chosen value has a large influence on the estimation of this particular parameter set. Hence, it could be misleading to to make claims about the physical meaning of $\alpha_l$ if the bounds are not properly defined. On the other hand, the original set led to estimates in the interior of the feasible parameter region and the bounds had no influence on them. As an extra disadvantage of this estimable parameter set, the sampling density of $\hat{\alpha}_l$ is truncated. Since it is positive on the feasible side, infinite on the constraint and zero on the unfeasible side, the computation of the covariance matrix becomes challenging \citep{bard:1974}.
\end{document}